\documentclass[journal]{IEEEtran}

\usepackage{graphicx} 

\title{Survey 3}

\author{
\IEEEauthorblockN{Riddhi Apte, Shubhada Gadgil, Gaurav S. Kasbekar, Rushikesh Patil, Prasanna Chaporkar}

\IEEEauthorblockA{
Department of Electrical Engineering \\
Indian Institute of Technology Bombay, \\
Powai, Mumbai - 400076}
}

\date{December 2025}

\usepackage[utf8]{inputenc}
\usepackage{cite}
\usepackage{graphicx}
\usepackage{amsmath,amssymb}
\usepackage{algorithmic}
\usepackage{array}
\usepackage{url}
\usepackage{multirow}
\usepackage{booktabs}
\usepackage{hyperref}
\usepackage{tabularx}
\usepackage{needspace}
\usepackage{float}
\usepackage{placeins}
\usepackage{kotex}
\usepackage{longtable}

\hypersetup{colorlinks=true, linkcolor=blue, citecolor=blue, urlcolor=blue}

\title{A Survey of UAV Communication Networks: Roles, Power Sources, and Security}

\begin{document}

\maketitle

\begin{abstract}
Driven by the demands of 5G/Beyond 5G and 6G networks, Unmanned Aerial Vehicles (UAVs) have surfaced in critical roles for aerial communications. In the present survey, we explore the multi-mode roles of UAVs as relays, User Equipment (UE), gNB and Reconfigurable Intelligent Surfaces (RIS), along with their deployment scenarios, architectural frameworks, and different communication models incorporating Artificial Intelligence (AI) configurations. We consider the effects of alternate power sources on the communication payload. The survey also aims to address security issues in UAV communications. As an advancement, we propose a novel UAV-Network-In-a-Box (NIB) architecture for disaster recovery and temporary coverage as an alternative to traditional network infrastructure.
\end{abstract}


\section{Introduction}
\IEEEPARstart{C}ommercial drone industry is embracing its growth in sectors like communications, surveillance, data collection, and defense \cite{ref1}. Strong demands for Unmanned Aerial Vehicle (UAV) enabled connectivity, aerial imaging (e.g., inspections of infrastructure), and expansion of autonomous UAV deployments have stimulated significant technological investments towards enhancing scalability and operational reliability. The accelerated approach if continued, by 2030, the Global Drone Industry is projected to surpass USD 160 billion driven by advancements in battery efficiency, use of lightweight composition of materials for design of UAV airframes (and components), AI-powered tasks and overall improvement in safety and their navigation systems \cite{ref1}. Further integration of UAVs with the latest communication technologies such as 5G/ 6G has enabled drones to be adopted as service models, making their entry in the Internet of Things (IoT) ecosystem a valuable trend as it enhances the real-time communication network and versatility for Beyond Visual Line of Sight (BVLOS) communications, making them suitable for commercial deployments \cite{ref1,ref2,ref3}.

As technological milestones are being achieved, UAV based applications have had a significant expansion in the communications domain like UAV-based Integrated Sensing and Communications (ISAC), where simultaneous environmental sensing and communications occur; they are expected to open up new horizons in wireless communications \cite{ref8}. Another recent milestone achieved between 2020 to 2025 is the introduction of 5G-enabled drones integrated with AI for surveillance of smart cities, where low-latency transmission of data is enabled through 5G networks to support real-time AI driven applications \cite{ref9}. In addition, UAVs have proven to be an essential component in public health emergencies like the COVID-19 pandemic, as reported by the United Children’s Fund (UNICEF), a rapid guidance report that used drones as a directional transport system for transportation of laboratory samples and other goods (medicines) significantly reduced human interaction thereby reducing exposure to infections \cite{ref10}. The applications discussed above depend on low-latency UAV communication links for command and control, real-time data transmission, and coordination with ground infrastructure, underscoring the central role of communications in enabling advanced UAV services.

In this survey, we primarily explore the roles of drones in communication areas, in particular their integration with 5G and beyond cellular networks as per dominant applications such as surveillance, signal relaying, IoT (smart cities), ad-hoc networking, network extenders for public communication networks and also in disasters and emergency recoveries \cite{ref11}. The system performance of such assisted communication networks is influenced by the UAV endurance and as most of the research and applications are based on battery-operated drones that offer limited flight endurance, there has been a marked rise in interest for the use of alternative power sources like wired power source to provide extended flight times (tethered UAVs) and fuel cells, for high-capacity operations. Renewable energy sources such as solar power and wind energy can boost environmental friendly and green operations. Thus, different power sources can support UAVs in relevant vital communication networks by providing stability, adaptability and reliability \cite{ref12,ref13}.

Apart from diversified power sources, UAVs vary based on the airframe, altitude of operation, range and the communicating network functions. All these factors have a direct influence on the suitability of UAVs for different applications. The classification of UAVs based on these parameters can be seen in Figure \ref{fig:1} below. For example, multi-rotor UAVs that operate at lower altitudes and cover shorter ranges (continuous power consumption for hovering) are suitable for surveillance, sensing and emergency communication where the UAVs act as a user equipment (UE) and relay. Fixed wing and hybrid UAVs offer an extension in endurance (they offer an aerodynamic lift) and range, making them reliable for relay, coverage, and backhaul. Other examples with representative use case scenarios can be seen in Table \ref{tab:1} \cite{ref14,ref15}. 
\begin{figure*}
    \centering
    \includegraphics[width=0.75\linewidth]{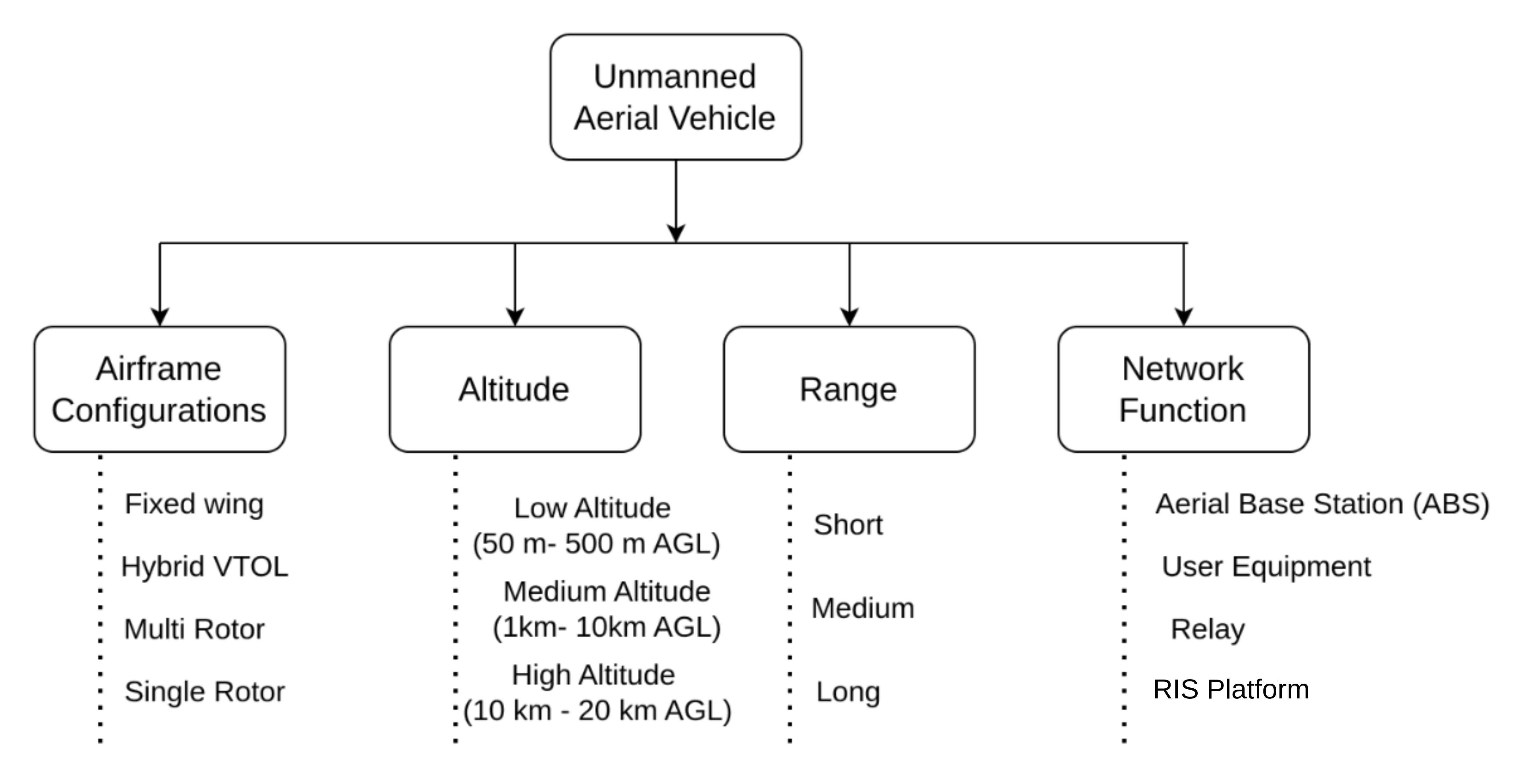}
    \caption{Classification of UAVs}
    \label{fig:1}
\end{figure*}

\begin{table*}[t]
\centering
\caption{Mapping of UAV Design Types to Communication-Oriented Use-Case Scenarios}
\label{tab:1}
\renewcommand{\arraystretch}{1.25}
\begin{tabular}{|p{2 cm}|p{3cm}|p{1.5cm}|p{1.5cm}|p{2.5 cm}|p{3.5 cm}|p{1.5 cm}|}
\hline
\textbf{Use-Case Scenario} & \textbf{Airframe Type} & \textbf{Altitude Class} & \textbf{Range Class} & \textbf{Network Function} & \textbf{Type of UAV Used and Why}  \\ \hline

Emergency \& disaster communication &
Multi-rotor / Hybrid VTOL &
LAP &
Short &
Aerial Base Station / Relay &
Hovering at low altitude lets rapid deployment in scenarios where terrestrial infrastructure is not available.  \\ \hline

Cellular capacity enhancement (hotspots) &
Multi-rotor / Hybrid VTOL &
LAP--MAP &
Short &
Aerial Base Station &
Due to higher altitude, line of sight (LoS) probability is increased, this helps in reuse of resources that enables capacity enhancement in congested areas.  \\ \hline

UAV-assisted relay communication &
Fixed-wing / Hybrid VTOL &
MAP &
Medium &
Relay Node &
High altitude relay positions increase coverage extension and link reliability between aerial and ground nodes. \\ \hline

BLoS communication &
Fixed-wing &
MAP--HAP &
Long &
Relay / Backhaul Node &
Long endurance and higher altitude enable extended communication links beyond terrain and infrastructure constraints.  \\ \hline

Tactical and military communication networks &
MALE UAV &
MAP &
Medium--Long &
Relay / Gateway &
Constant airborne relays support, adaptive communication networks. \\ \hline

IoT data collection \& aggregation &
Multi-rotor / Fixed-wing &
LAP--MAP &
Short--Medium &
Gateway / Relay &
UAVs function as mobile data collection, aggregating data from spatially distributed IoT and sensor nodes.\\ \hline

Surveillance \& infrastructure monitoring &
LALE UAV &
LAP &
Short &
User Equipment / Relay &
Reliable connectivity is required for continuous transmission of sensing and monitoring data. \\ \hline

Wide-area connectivity / NTN integration &
HALE UAV &
HAP &
Long &
Gateway / Backhaul &
High-altitude platforms provide wide-area LoS coverage, enabling gateway functionality analogous to non-terrestrial networks. \\ \hline

\end{tabular}
\end{table*}

This survey also aims to review the current trends and challenges associated with the diverse roles of UAVs in communication networks, integration of evolving technologies with UAVs, the effects of different power sources on communication payloads, and physical-layer security in UAV communication modules.

\subsection{Objectives and Contributions}
UAVs in the communications field have gained remarkable advancement in communication technologies, integration with 5G and beyond cellular networks, reflecting a surge of academic research and publications, emphasizing their crucial roles in communication systems. Significant efforts have been made towards integrating existing communication technologies with emerging innovations, such as Artificial Intelligence (AI) and swarm coordination, to enhance UAV network performance and operational efficiency. Despite this, most existing surveys focus only on a subset of UAV functionalities such as UAV as a UE, UAV as a relay, or UAV as an aerial communication node, without providing a broader perspective on the ecosystem of UAV-enabled networks, including multi-role operations, power management, and autonomous deployment.
For instance, \cite{S1} provides an analysis of UAVs enabled ultra-reliable low-latency communications (URLLC), that emphasize on performance factors of latency and reliability in applications (military and defense), while \cite{S2} describes the roles of UAVs in wireless communications, based on technological challenges, deployment scenarios, and practical applications (smart agriculture, data offloading, event based coverage). The survey \cite{S3} focuses on network security and covers vulnerabilities, threats, and mitigation models applicable in UAV networks. The authors of \cite{S4} explore different coordination strategies, the sharing of resources, and cooperative communications in UAV based communication networks. In \cite{S5},  a comprehensive survey of UAV-assisted data collection in wireless sensor networks has been presented, based on factors such as coverage, routing, and scheduling. The survey in \cite{S6} addresses fundamental issues such as architecture, mobility management, and reliability in UAV communication networks. The survey \cite{S7} discusses practical aspects of deployment, regulations, and standardization of UAVs in cellular systems, whereas \cite{S8} further highlights the applications and challenges based on coverage enhancement and communication performance for UAVs in cellular systems. Also, \cite{S9} describes UAV applications, open research problems, and design considerations, along with trajectory optimization, coverage maximization, and open research challenges like A2G channel modeling, interference, and practical deployment. The survey in \cite{S10} focuses on UAV integration with 5G-and-beyond networks, discusses emerging technologies, regulatory aspects, and research trends, and \cite{S11} examines 5G network slicing in UAV-assisted communication networks, while describing use cases and future trends in such networks.
Thus, this survey primarily tries to fill a gap present in other surveys and other identified blind spots mentioned below:
1) Not all UAV roles (UE, relay, gNB, and RIS) have been collectively discussed,
2) Previous surveys have passed over the effects of power systems on the communication payload,
3) UAV-assisted digital twinning in wireless networks,
4) UAV network security in brief,
5) UAV as a Network-In-a-Box (NIB) that is capable of establishing a standalone network.
This survey aims to address these gaps and provide a very comprehensive perspective on UAVs and their role in communication networks by incorporating various aerial nodes
and their potential.

\subsection{Paper Structure}
The remainder of the paper is structured as follows. Section I discusses the evolution of standards relevant to UAVs. Section II discusses integration of UAV with emerging technologies. Section III states the different roles of UAVs in communication networks. Section IV examines the effects of different power sources on communication payloads. Section V provides a brief overview of security  considerations in UAVs along with physical layer vulnerabilities and their mitigation. Section VI briefly describes an architecture that is designed to overcome the limitations of traditional frameworks and Section VIII briefs the future research directions before concluding the paper.  

\begin{figure*}
    \centering
    \includegraphics[width=0.75\linewidth]{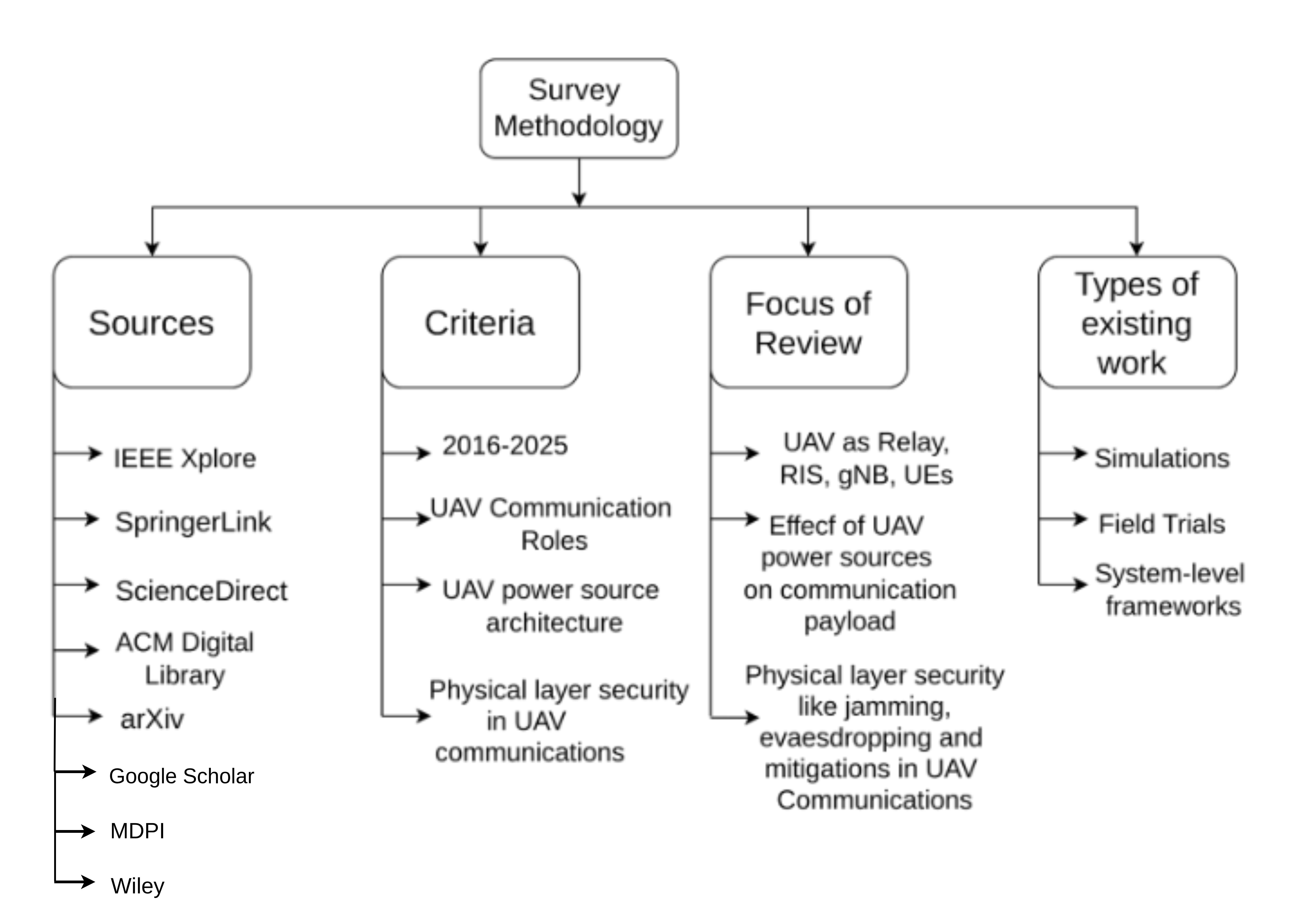}
    \caption{Methodology flowchart}
    \label{fig:2}
\end{figure*}

The survey flowchart is illustrated in Figure \ref{fig:2}. For a clear overview and coverage of this scope, refer to Table \ref{tab:2}. It highlights the number of literature papers analyzed, UAV power source architectures and different roles of UAVs in communications, along with the evaluation models, simulation studies and real-world test methods.

\begin{table*}[t]
\centering
\caption{Survey Metrics and Key Observations}
\label{tab:2}
\renewcommand{\arraystretch}{1.2}
\begin{tabular}{|p{4.5cm}|p{8.0cm}|}
\hline
\textbf{Survey Metric} & \textbf{Remark} \\ \hline

Total Papers Reviewed &
169 \\ \hline

Years Covered &
2016--2026 \\ \hline

Sources Used &
IEEE Xplore, ScienceDirect, SpringerLink, ACM, arXiv \\ \hline

UAV as Communication nodes &
Relay, UE, Aerial gNB, RIS-mounted \\ \hline

Power Systems &
Li-Po Battery, Renewable Energy, Fuel Cell, Tethered, Hybrid \\ \hline

Simulation Studies and Real-World Testbeds &
25 + \\ \hline

Evaluation Models &
Friis Equation, Path-loss, SNR, Kalman Filtering, Convex Optimization, Dinkelbach Method, MATLAB/Simulink, LoS/NLoS \\ \hline

\end{tabular}
\end{table*}

\section{Evolution of standards for UAV communications}
The integration of UAVs with wireless networks has prompted standardization bodies to define protocols and standards that ensure their reliability and efficient operations. 3GPP has developed standards for LTE support in UAVs to address the issues faced by UAVs at high-altitude operations (approximately 300 meters). Further 3GPP formally defines an UAV as only an aerial UE and the Command and Control Link (C2 Link) (C2 Link is used to transmit commands from ground control station to UAVs and feedback from telemetry)  signals to the aerial UEs are established while also focusing on UAV traffic management and air to ground (A2G) communication channels \cite{ref16}. 

The IMT-2020 (5G framework) highlights UAVs’ vital role in 5G applications such as eMBB (enhanced mobile broadband) and mMTC (massive machine type communication), which also support AI operations, 3D coverage, and connectivity. IMT-2030 (6G) focuses on 6G and expands it towards ISAC and a growing omnipresence of aerial and terrestrial network integration \cite{ref17,ref18}.

ETSI GR ENI (Group Report – Experiential Networked Intelligence) focuses on  a 6G oriented framework that highlights the role of AI in self optimizing communication networks that can adapt to different dynamics depending on varying operational conditions. While it is not an UAV specific standard, the principles of these adaptations can be applied to UAV supported communication networks where any autonomous resource management is required. ETSI TC SES (ETSI Technical Committee on Satellite Earth Stations and Systems) focuses on interoperability among satellite and other aerial nodes, even though they do not introduce any UAV-assisted/ specific communication procedures, but they establish the requirements for beyond line of sight (BLOS) communication that shall let UAVs operate autonomously \cite{ref19,ref20}.

The IEEE like 3GPP and ETSI have developed multiple standards that support evolution of UAVs ecosystems, addressing networking, communication, payload integration, traffic management, and power interfaces.
\begin{enumerate}
    \item IEEE Std 1930.1-2022 presents a software-defined networking (SDN) framework specifically designed for aerial networks, promoting centralized control and programmability that are critical for managing the complexity of UAV operations.
    \item IEEE Std P1920.1 provides guidelines for aerial communications, detailing interface definitions and network functionalities that may facilitate coordinated multi-UAV operations in mission-critical scenarios.
    \item IEEE 2874-2025 defined Hyperspace Modeling Language (HSML), a data language that lets machines like UAVs describe things clearly to operators and other machines like UAVs with Hyperspace Transaction Protocol (HSTP). The HSTP protocol enables machines like UAVs to exchange and act on the information based on HSML \cite{ref21,ref22,ref23}.
    \item IEEE 1937.1-2020 defines standards for UAV payloads and specifies requirements for their environmental robustness and minimal functionality that includes tolerance to temperature extremes, moisture, and resistance to corrosion \cite{ieee1}.
    \item IEEE P1936.1-2021 describes standards focusing on UAV application scenarios and parameters that are required for execution of operations (platform characteristics, safety requirements and operational limit parameters for a lightweight UAV) \cite{ieee2}. 
    \item IEEE P1939.1-2021 standardizes supervision for low- level UAV traffic by defining routing, sensing and UAV Traffic Management (UTM) communication protocols \cite{ieee3}.
    \item IEEE P1920.1-2022 illustrates processes for self organizing A2A ad hoc networks supporting wireless technologies by defining data formats, service architectures, and privacy frameworks \cite{ieee4}.
    \item IEEE P1954 defines an architecture that allows UAVs to switch in between frequencies based on availability and crowding \cite{ieee5,ieee6}.
    \item IEEE P1937.9 defines standards related to wired/ non-wired power source interfaces to UAV platforms \cite{ieee7}.
    \item IEEE 1937.8-2024 specifies standards related to interfaces, functionality, reliability, and safety requirements for cellular communication terminals mounted on UAVs \cite{ieee7, ieee8}.
\end{enumerate}

\begin{figure*}
    \centering
    \includegraphics[width=1.0\linewidth]{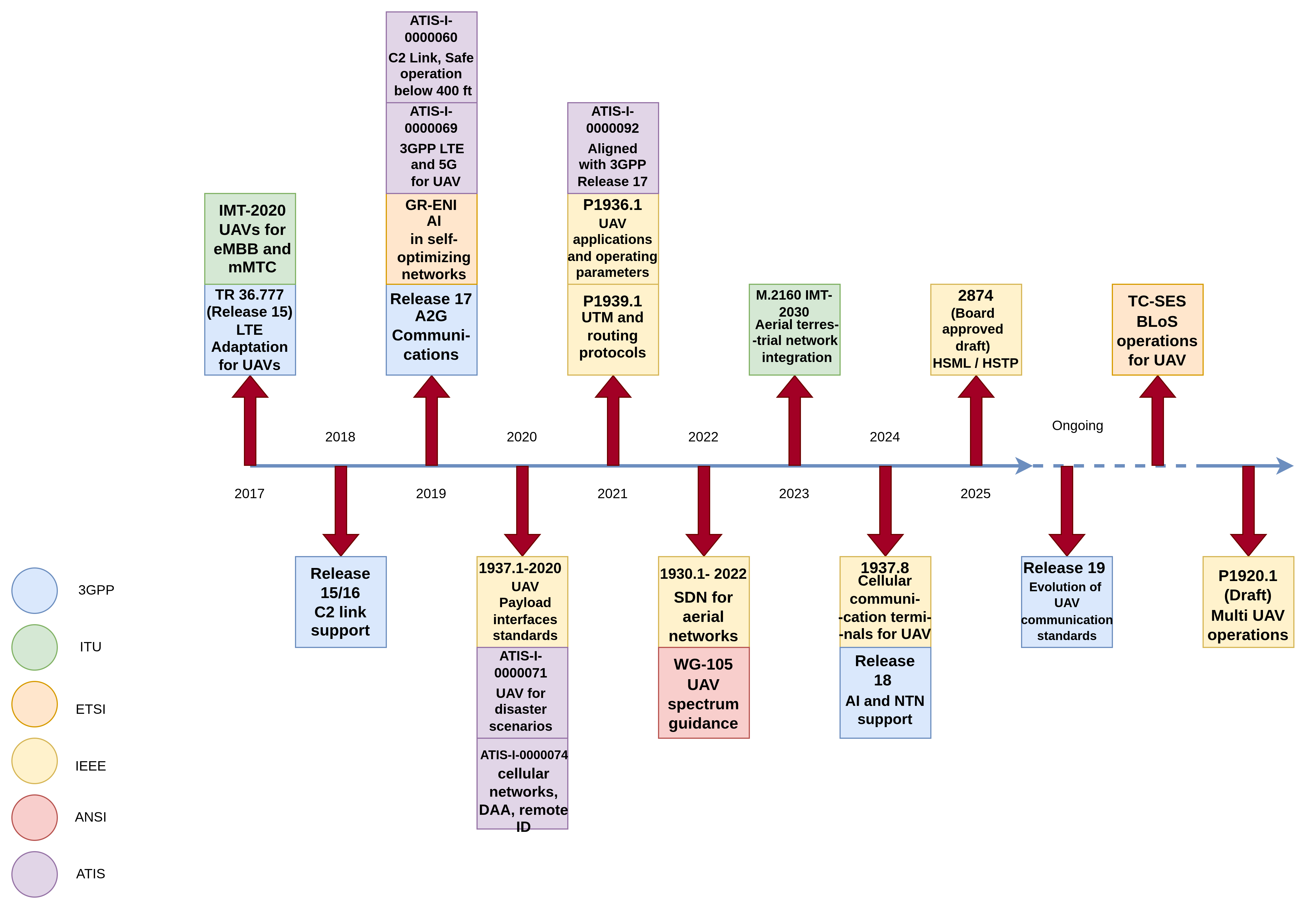}
    \caption{Evolution of standards}
    \label{fig:3}
\end{figure*}

Further elaborative summaries on standards for UAVs are indicated in Table \ref{tab:3} and the evolution of the standards for UAVs is illustrated in Figure \ref{fig:3}.

\begin{table*}[!t]
\centering
\caption{Evolution of Standards}
\label{tab:3}
\renewcommand{\arraystretch}{1.2}
\begin{tabular}{|p{2cm}|p{3cm}|p{2cm}|p{2cm}|p{5 cm}|}
\hline
\textbf{Standard Body} & \textbf{Standard Name} & \textbf{Release/Version} & \textbf{Year of Release} & \textbf{Relevance / Core Focus of the Standard} \\ \hline

3GPP & TR 36.777 & Release 15 & 2017 & Addresses mobility and interference issues for User Equipment (UE) operating at high altitudes (up to approx 300 m) within LTE networks but can be extended to other networks. \\ \hline

3GPP & C2 Link support & Release 15/16 & 2018 & Defines support for highly reliable, bidirectional Command and Control (C2) Link signals (commands/telemetry feedback) for specific device types. \\ \hline

3GPP & A2G Communication \& Traffic Management & Release 17 & 2019 & Focuses on optimizing communication channels and protocols for Air-to-Ground (A2G) scenarios and traffic management of mobile nodes. \\ \hline

ITU & Report M.2410 – IMT-2020 (5G Framework) & – & 2017 & Defines the 5G framework and highlights the role of systems in applications such as eMBB, mMTC, and supporting AI operations and 3D coverage. \\ \hline

ITU & Rec. M.2160 – IMT-2030 (6G Framework) & First Version & 2023 & Defines critical elements in 6G systems, focusing on Integrated Sensing and Communication (ISAC) and the integration of atmospheric and terrestrial networks. \\ \hline

ETSI & GR ENI – Experiential Networked Intelligence & v1.1.1 & 2019 & A 6G-oriented framework for using AI in self-optimizing communication networks that adapt to varying operational conditions and dynamics. \\ \hline

ETSI & TC SES – BLOS Operability & Draft & Ongoing & Focuses on interoperability among satellite and other high-altitude communication nodes, establishing requirements for BLOS communication. \\ \hline

IEEE & IEEE 1930.1 – SDN for Aerial Networks & Std 1930.1-2022 & 2022 & Presents a Software-Defined Networking (SDN) framework specifically for aerial networks, promoting centralized control and programmability. \\ \hline

IEEE & IEEE P1920.1 – Aerial Communication \& Networking & Draft & Ongoing & Provides guidelines, interface definitions, and network functionalities to facilitate coordinated operations in mission-critical aerial scenarios. \\ \hline

IEEE & IEEE 2874 – Ethical Coordination & Board Approved Draft & 2025 & Establishes the Hyperspace Modeling Language (HSML) for machine-readable descriptions and the Hyperspace Transaction Protocol (HSTP) for data exchange and action among automated machines. \\ \hline

ANSI & WG-105 UAS Integration & Guidance / Pre standard & 2022 & Guidance on UAV communication (C2 management, spectrum, communication security, use of Dept of defense defined and non aviation frequency band for UAV operations \cite{ansi1}. \\ \hline

ATIS & ATIS-I-0000060 & Issue I-0000060 & - & States connectivity for UAV using C2 link, QoS, safe operations below 400 ft \cite{atis1} \\ \hline

ATIS & ATIS-I-0000069 & Issue I-0000069 & - & Analysis of 3GPP LTE and 5G evolution for UAV \cite{atis2} \\ \hline

ATIS & ATIS-I-0000071 & Issue I-0000071 & - & States guidelines for restoring UAV communication during scenarios like disasters, spectrum use and operational usage \cite{atis3} \\ \hline

ATIS & ATIS-I-0000074 & Issue I-0000074 & - & Integration of UAV operations with cellular network, DAA and remote ID \cite{atis4} \\ \hline

ATIS & ATIS-I-0000092 & Issue I-0000092 & - & 3GPP release 17 supports UAV operation and enhancement for industrial and recreational UAV use cases \cite{atis5} \\ \hline
\end{tabular}
\end{table*}

\subsection{Sample Simulation Parameters for UAV Assisted Aerial Base Station Deployments} 
The standards and guidelines summarized in the above section and Figure 3 establish the operational and regulatory framework required for UAV communication networks. To be able to use/ apply these standards in real-world scenarios, sample simulation parameters that are derived from 3GPP specifications are provided as a base for evaluation of  UAV mobility, characteristics, and network performance and can be found in Table \ref{tab:4} \cite{ref16,samp1,samp2}.

\begin{table*}[htbp]
\centering
\caption{System Parameters for Urban Macro, Rural Macro, and Disaster Scenarios}
\label{tab:4}
\begin{tabular}{|p{2cm}|p{3cm}|p{3cm}|p{3cm}|}
\hline
\textbf{Parameter} & \textbf{Urban Macro (UMa)} & \textbf{Rural Macro (RMa)} & \textbf{Disaster} \\
\hline
UAV Height & 50-120 m (AGL) & 80 m (AGL) & 100-120 m (AGL) \\
\hline
Cell Layout & Hexagonal (500 m) & Hexagonal (1732-5000 m) & On-demand \\
\hline
Active UEs & 32-64 & 10-20 & 20- 100 \\
\hline
Carrier Frequency & 2-60 GHz & 700 MHz (optionally 800 MHz) & 2-6 GHz; 12-18 GHz (Ku); 20-30 GHz (mmWave) \\
\hline
System Bandwidth & 10 MHz & 10 MHz & 20- 100 MHz \\
\hline
Total BS Transmission Power & 46/49 dB (10/20 MHz) & 46/49 dB (10/20 MHz) & 43 dB \\
\hline
Base Station Noise Figure & 5 dB & 5 dB & 5-9 dB \\
\hline
Terrestrial UE Tx Power & 23 dB & 23 dB & 23 dB \\
\hline
Aerial UE Tx Power & 23 dB & 23 dB & 23- 26 dB \\
\hline
Mobility & Pedestrian / Vehicular & Pedestrian / Vehicular & Pedestrian / Vehicular \\
\hline
UE Speed & UE: 30 km/h; Indoor: 3 km/h; Aerial: 160 km/h & UE: 30 km/h; Indoor: 3 km/h; Aerial: 160 km/h & 0-50 km/h \\
\hline
UAV Speed & Max. 160 km/h & Max. 160 km/h & Max. 160 km/h \\
\hline
\end{tabular}
\end{table*}

\section{UAV integration with Emerging technologies}
\begin{figure*}
    \centering
    \includegraphics[width=1.0\linewidth]{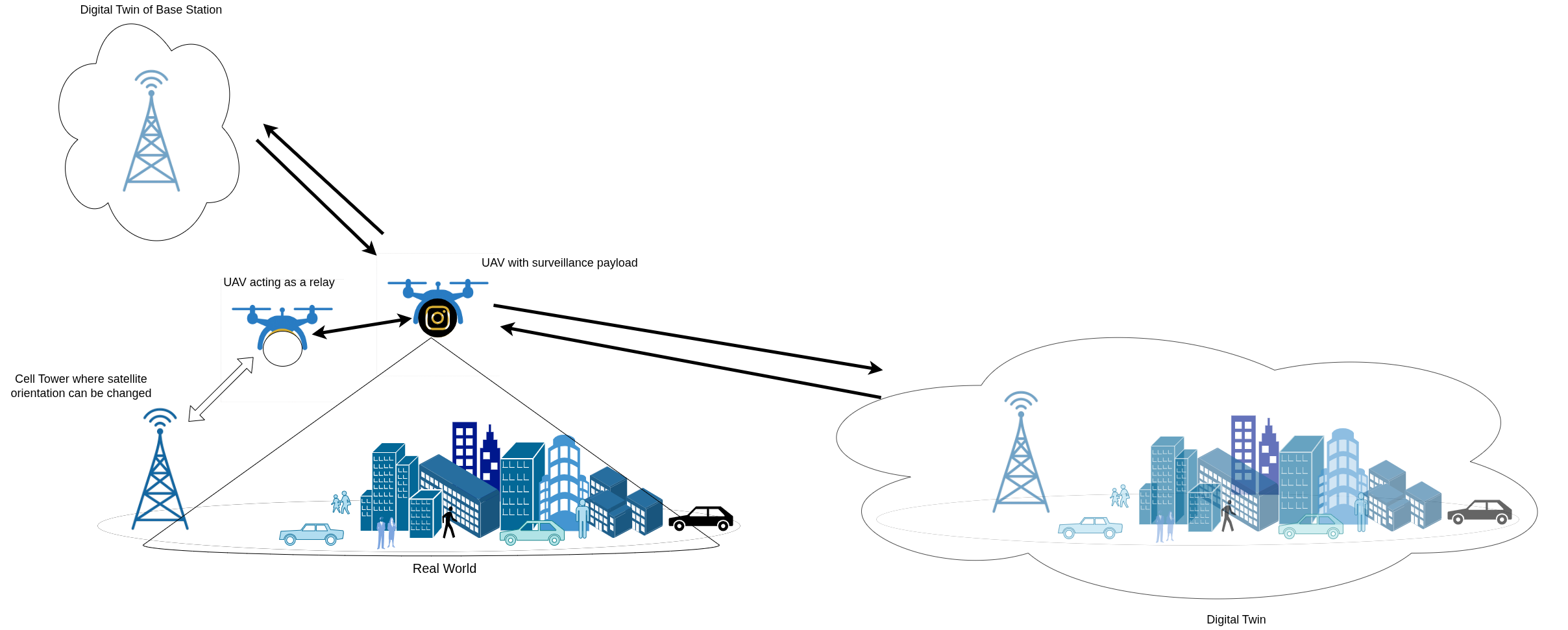}
    \caption{Digital twinning using UAVs}
    \label{fig:4}
\end{figure*}

As communication networks approach beyond 5G and towards 6G, UAVs are paving their way to become smarter and faster by integrating with emerging technologies. One promising approach is the integration of digital twin (DT) technology, to create a virtual replica of an UAV based communications network either of an aerial base station or a relay network as illustrated in Figure \ref{fig:4}, which illustrates a bidirectional (two-way) sync environment. A UAV collects data from a base station in the physical environment and forwards it to a cloud DT server for storage. The DT framework processes and updates the virtual model to support monitoring, predicting, and optimizing the actual network. The control and optimization data is fed to the UAV via the virtual network scenarios making it a closed loop for management of the network. The virtual model is continuously updated with real time information which can be obtained from real time telemetry and performance parameters of the physical network that enable simulation, monitoring and optimization of the physical network are updated in real time. By utilizing the DT technology, UAV networks can achieve goals like enhanced coverage, monitoring wear and tear for timely replacement of components and proper allocation of resources, which will increase the reliability due to more predictability in environments, deployment scenarios of the infrastructure, user mobility and network topology \cite{ref24,dt1,dt2}.
ISAC is gaining popularity, where UAVs use a combined wave for both communications (with ground stations, terrestrial UEs, and other UAVs) and sensing the environment (detecting objects and observing terrains). Through ISAC, the simultaneous integration helps UAVs to perform communication and sensing over the same spectrum. This makes efficient spectrum resource management possible and spectrum resources are also conserved. This simultaneous approach is also useful in search and rescue operations \cite{ref25,et1,et2}.
Another trend that can be focused on is the integration of UAVs with AI. Using AI, dynamic allocation of spectrum, trajectory planning, swarm formation, and UAV channel selection requirements can be amended as per the operation whereabouts \cite{ref26,ai1,ai2}.

One more emerging technology that is being considered is UAV-enabled Internet of Battlefield Things (IoBT), an extension of IoT that interconnects battlefields using UAVs as it combines newer capabilities— such as UAV networking, edge computing, AI, and 5G/6G tactical communications— to create intelligent and adaptive battlefield networks \cite{et3,et4}

These integrations will enhance the role of UAVs in cross-domain dynamics and further make the UAV nodes intelligent in communication networks.

\section{UAV roles in communications}
Ongoing developments in wireless networking are progressively integrating UAVs into 5G networks as reconfigurable nodes. These platforms have evolved beyond their flying capabilities and serve as a relay function, gNB, UEs, and RIS. UAVs offer enhanced connectivity, path planning and rapid deployment for the extension of the network (lightweight as compared to traditional networks). This section systematically highlights these roles. A comparative Table \ref{tab:5} further elaborates shortcomings and features of the UAV-enabled network against the traditional network infrastructure. 

\begin{table*}[t]
\centering
\caption{UAV Enabled and Traditional Ground Networks}
\label{tab:5}
\renewcommand{\arraystretch}{1.2}
\begin{tabular}{|p{4.5cm}|p{5.5cm}|p{5.5cm}|}
\hline
\textbf{Aspect} & \textbf{UAV-Enabled Networks} & \textbf{Traditional Ground Networks} \\ \hline

Deployment & Rapid deployment in areas & Fixed structure with limited scalability \\ \hline

Energy Constraints & Requires energy management & Energy constraints are less stringent as compared to UAV enabled networks \\ \hline

Interference and Mobility & Prone to interference; dynamic mobility & Stable handovers and mature interference coordination \\ \hline

Security & Susceptible to jamming, spoofing, and eavesdropping & Better protected via physical security and regulated environments \\ \hline

\end{tabular}
\end{table*}
\subsection{UAV as an Aerial Base Station}
\begin{figure*}
    \centering
    \includegraphics[width=0.75\linewidth]{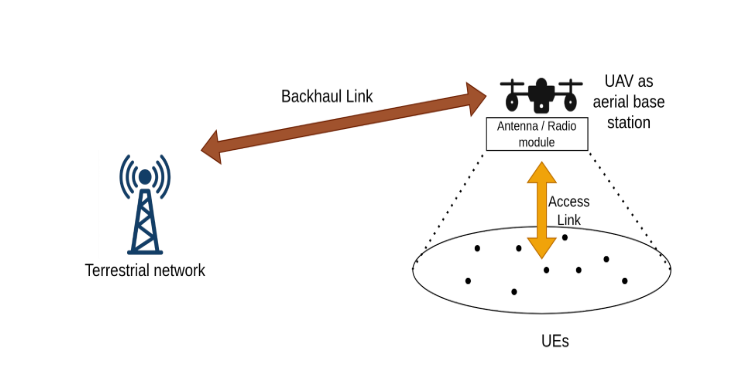}
    \caption{UAV enabled aerial base station}
    \label{fig:uavbs}
\end{figure*}

The Next Generation Node Base Station (gNodeB), also known as gNB, is a crucial component of 5G technology. The primary role of gNB is to propagate a signal to User Equipment (UE), which involves modulating/demodulating the signal as well as antenna technology. It also manages the allocation of radio resources to the UEs. UAVs as gNB carry a payload comprising of 5G New Radio (NR) units, Distributed Units (DU)/ Control Unit (CU), antennas to act as aerial base stations to provide temporary coverage, flexible deployment, LoS communications, capacity management and provide cost-effective solutions \cite{ref27}. Figure \ref{fig:uavbs} illustrates a UAV-enabled aerial base station, where a drone provides wireless connectivity to the UEs. This setup consists of two primary wireless links, one of which connects the UAV to the core network (backhaul) and the other that connects the UAV to users (access) to provide coverage.

UAVs can provide rapid coverage extension by repositioning themselves in 3D space to serve high-demanding and hard to reach areas. They also offer greater flexibility in resource management, as transmission power, spectrum allocation, and antenna beam directions can be dynamically adjusted based on user density, mobility, and air-to-ground channel conditions. This capability makes UAVs more adaptable than traditional fixed networks, particularly in scenarios with high user mobility. Unlike traditional networks, UAV based networks face challenges like UAV mobility, limited flight endurance (power), dynamic A2G and A2A channels, high user mobility, coverage optimization, and interference management in UAV networks. To address this, the authors in \cite{ref28} propose a Joint Scheduling and Resource Allocation (JSRA) algorithm, where two components user scheduling (determines which users should be served at any given time) and radio resource allocation (power, bandwidth block assignment) are implemented to work together. These tasks are performed by the UAV that acts as an aerial base station and UEs are on the ground. This is achieved by employing an Advantage Actor Critic Reinforcement Learning (A2C RL) that can adapt dynamic scheduling and resource allocation based on user density, mobility, and A2G channel conditions. Simulations for realistic environments were carried out using this RL method. Improvements in throughput, reduction in packet drop rate, and increased fairness of use (sharing of network resources) compared to static allocation methods were achieved.

However, the RL based approaches require a complex training module which should have high computational capacity to support frequent updates of the RL. This increases the need for extensive real-time training data. To address these limitations in specific scenarios (e.g., post disaster), the authors in \cite{ref29} introduce a traffic control strategy that configures a gNB based on narrowband Internet of Things (NB-IoT) principles for Low Power Wide Area Networks (LPWANs). This approach is specifically designed for UAV-assisted IoT applications involving small data transmissions, extended battery life, and wide-area coverage, rather than general high-throughput communication services. Unlike the RL-based scheduling, the proposed deterministic scheduling framework avoids stochastic decision-making, making UAV traffic more predictable and easier to manage. The algorithm acquires channel parameters such as line-of-sight conditions and link quality, identifies service types, applies beamforming, and performs resource optimization to maintain quality-of-service (QoS). This framework is further integrated with millimeter wave (mmWave) and massive multiple-input and multiple-output (MIMO) technologies to enhance coverage. UAV coverage zones are also formulated using graph theory, which is used as a modeling and optimization tool. Simulation-based evaluations are conducted to assess the throughput performance.

\begin{figure*}
    \centering
    \includegraphics[width=0.75\linewidth]{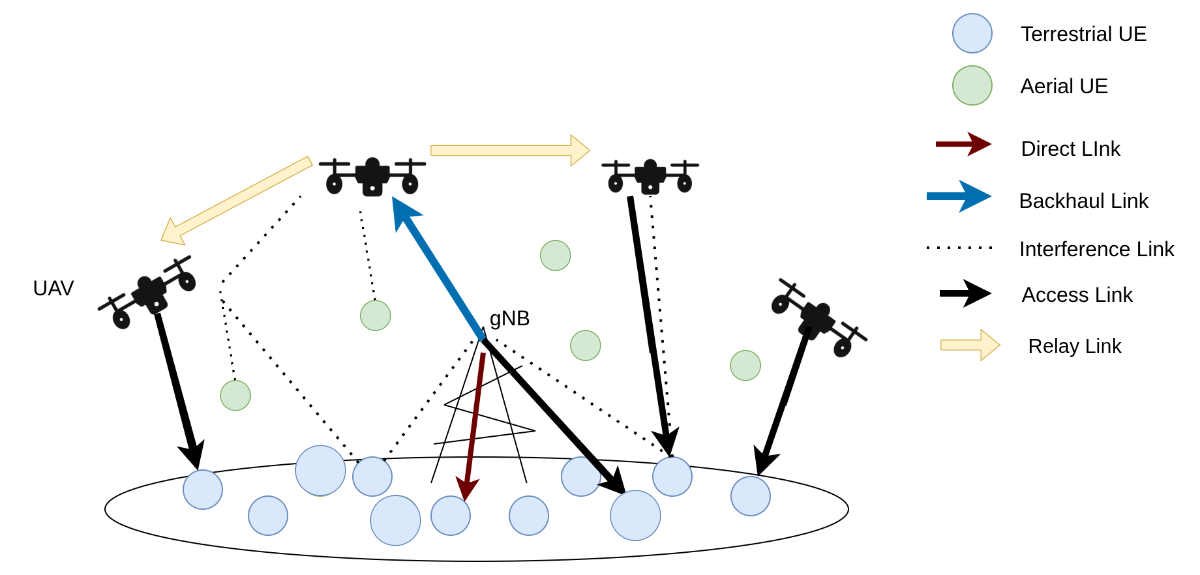}
    \caption{Multilevel UAV IAB setup}
    \label{fig:5}
\end{figure*}

The above NB-IoT approach raises several concerns, as it is designed for low-rate, low-power IoT devices and hence, unsuitable for higher throughput communication links like that of UAVs. Consequently, it is limited to low data rate signals rather than broadband aerial access. When UAVs act as aerial base stations, they can be efficiently deployed in high-mobility scenarios such as highways to improve cellular handover performance. Deploying traditional base stations in such dynamic and temporary environments is costly, whereas UAVs can serve as flexible, temporary base stations. To address UE handovers between terrestrial (fixed) gNB and UAV-mounted gNBs, the authors of \cite{ref30} propose a machine learning–based framework. The proposed approach is as follows: 
\begin{enumerate}
    \item Online real-time data is captured and compared with pre-trained data to ensure accurate handovers while minimizing delays and ping-pong rates (the rate at which UAVs switch between base stations/ cells). 
    \item Offline training data is provided and cleaned using Kalman filtering (a mathematical tool to predict the state of a system even if the data present is noisy) to eliminate noise and fading.
\end{enumerate}
Despite its promising performance, this approach assumes the availability of accurate real-time and pre-determined network data along with sufficient onboard computational infrastructure, which may not always be feasible in a dynamic UAV environment. The reasons being mobility, limited energy availability, weather variations, and airspace constraints (regulatory restriction, flying altitude, safety).
Their simulation results show significantly reduced handover delays (21\% to 1.2\%). Inspite of this, the network can still face challenges due to a sudden increase in network congestion and abrupt geographical terrains. The approach is also assumed to always have access to accurate, real-time, and pre-determined data, which is not always possible in a dynamic environment like that of UAVs \cite{ref30}.
Most of the papers on UAVs as aerial base stations focus on UE to UAV  handovers, without addressing the backhaul resource allocation, interference between UAVs, or multi-level UAV scenarios. Thus, one more role of UAVs as aerial base stations is in Integrated Access and Backhaul (IAB) scenarios. For example, this is the case in \cite{ref31}, where UAVs act as aerial base stations and provide both access and backhaul links, and by following this, spectral resources are allocated to both access and backhaul. Using this, traffic demands can be efficiently addressed. A multi-level approach can be considered as illustrated in Figure \ref{fig:5} that uses a ground gNB to provide wireless backhaul links to UAVs that relay traffic to other components through multi-hop backhaul connections. These UAVs simultaneously establish access links to both terrestrial and aerial UEs located within the coverage region. Direct links, access links, and backhaul links coexist and reuse the same spectral resources. To use the same spectral bands, frequency division duplex (FDD) can be used; other ways can be integrating massive MIMO to maximize bandwidth reuse. Along with these factors, other aspects like 3D positioning of UAVs, power consumption, and precoder designs also play an important role. The authors of \cite{ref31} consider two scenarios for simulations:
\begin{enumerate}
    \item Scenario A considers multiple groups of UEs in different hotspots. An algorithm prioritizes connections to the nearest base station, reducing latency and improving throughput.
    \item Scenario B considers a single group of UEs in a hotspot, where increasing the UAV altitude shows improvements in coverage and capacity. 
\end{enumerate}

The simulation results indicate that IAB-based UAVs can improve the throughput, offer better network coverage, and increase data rates for better QoS. However, due to the multi-level arrangement, UAVs can interfere with different layers (access layer, backhaul layer, or multi-hop layer).

UAV-based IAB faces limitations like restricted deployment facilities and power constraints, vulnerability to blockage due to obstacles such as buildings, unknown geographical terrains, and signal blockages by swarm of UAVs. These disruptions can decrease the reliability on the channel link that require strong LoS and dependence on mmWave communications. These issues can be mitigated using a mmWave Distributed Phase Array (DPA) architecture. The DPA design can avoid LoS blocking and heating issues in the architecture. DPA improves robustness against blockage through spatial diversity, distributed antenna placement, and adaptive beamforming, allowing alternate propagation paths to be exploited when the dominant LoS link is obstructed. The authors of \cite{ref32} propose a 4-layer DPA system where independent antenna beams are generated using hybrid beamforming, allowing for precise spatial multiplexing and reduced interference, incorporating a split for further enhancing thermal efficiency. Splitting the radio frequency (RF) signal into 2 or more intermediate bands also makes DPA systems less prone to signal degradation than IAB setups. If the antennas in DPA systems are elaborated with ceramic materials, it gives superior thermal connectivity and reduces the risk of performance drops in high-temperature conditions.

Other studies are elaborated in Table \ref{tab:bs}. 
\begin{table*}[ht]
\centering
\caption{Comparative Analysis of UAV Base Station and Network Studies}
\label{tab:bs}
\scriptsize
\setlength{\tabcolsep}{2pt}
\renewcommand{\arraystretch}{1.0}
\begin{tabularx}{\textwidth}{|p{0.9cm}|X|p{3.2cm}|p{3.2cm}|}
\hline
\textbf{Ref.}  &  \textbf{Summary} & \textbf{Advantages} & \textbf{Limitations}\\
\hline
\cite{base1}  & Proposes a base station selection and deployment algorithm using A- Star Algorithm (provides shortest path between source and destination) with an aim to reduce switching overhead, it also helps in reducing energy consumption and complexity in operations & Shortest path achieved and reduced energy consumption & Pre-planned, can struggle if obstacle placements is changed\\
\hline
 \cite{base2} & This paper focuses on edge prior placement algorithm in which UAVs detect users which are not close enough to the base station creating a covering circle around these users based on UAVs capacity. It calculates the ideal 3D height for signal strength, removes the covered users, and repeats the process inward until the entire area is connected & Number of UAVs required is minimized, prioritizes the users & Ignores user density at the center of the cell and may also ignore SNR\\
\hline
 \cite{base3} & This paper focuses on 3-D mapping, probability of maximizing LoS connectivity of UAV with ground infrastructure, model optimizes UAVs position based on real time movement of the users.   & Tracks users and optically places UAV along with the users, LoS is maintained & High speed processing, constant tracking increases power consumption\\
\hline
 \cite{base4} & This paper states the use of HAPs as an intermediary platform that can offer low latency and wide area coverage. This can help gap the high latency caused due to SATCOMM. These platforms are the key to providing internet to remote oceans and mountainous regions where ground infrastructure is impossible to build. & Single stratospheric UAV can cover an area up to 200 km & Must survive extreme environmental conditions and high latency\\
\hline
 \cite{base5} & This paper explores the optimization of the hardware (MIMO, C2 hardware to control flight, network slicing support) that UAV-BS that can handle 5G protocols. The authors focus on a small cell (sports stadium) concept where UAV- BS will act as mobile communication node to handle temporary spikes in data traffic & Provides fiber like speeds, priority to emergency traffic & Heavy MIMO antennas, mmWaves can experience attenuation\\
\hline
\cite{base6} & This paper analysis 3D UAV model (3D Poisson Point Process with a focus on Downlink. Using probability and stochastic geometry, maximum number of users an UAV (Random waypoint to determine how UAVs move between users) can support can be determined before the signal quality degrades & Prediction of exact users & Poisson Point Process (PPP) are idealized \\
\hline
\cite{base7} &  The paper states a deployment method based on Artificial Bee Colony (ABC) algorithm, a swarm intelligence algorithm  that mimics scouting techniques of bees for food, where the areas requiring signal coverage assistance as sources and UAVs finding optimal positions to maximize signal coverage for users & Can cover huge areas & Complexity increases with number of users  \\
\hline
\cite{base8} & Using a physical test bed and a two tier UAV system where higher layer of UAV provides backhaul to lower layer that collects data using sensor payload for IoT proves that network can be extended for several miles & Inexpensive technology and power limited sensors & Not suitable for video/ voice and ideal for small data\\
\hline
 \cite{base9} &  Based on actual flight trials in London that were used to measure how a standard (LTE) Signal traveled from ground towers to UAV, based on measured factors like signal strength and quality, it was concluded that up-tilt (Vertical angle of antenna) interference from ground towers is a challenge for the flying UAVs & Standard LTE equipment can support UAVs, existing cellular infrastructure can enable Beyond-Visual-Line-of-Sight (BVLOS) flights at a low cost & UAVs pick noise from multiple towers, leading to high noise levels\\
\hline
 \cite{base10} & This paper aims at placement and resource management and proposed an algorithm which that states UAVs must decide the division of slice of the available radio resources along with trajectory optimization for high coverage & UAVs power and position is optimized to keep the signal stable & If drone moves slowly, then loses track of users \\
\hline
\cite{base11} &  This paper focuses on DRL algorithm for managing of content delivery in IoT networks. UAVs in these are caching and also delivering content related to software updates, DRL agent learns through trial and error which flight paths and transmission schedules result in the highest efficiency and lowest energy consumption & No manual programming & Requires a long training data and can fail to predict in emergency scenarios\\
\hline
\end{tabularx}
\label{tab:UAV-BS-summary}
\end{table*}

Thus, UAVs as gNBs can be integrated with different solutions to make the network more robust while keeping it flexible during deployment.

\subsection{UAV as a Relay Node}
\begin{figure*}
    \centering
    \includegraphics[width=0.75\linewidth]{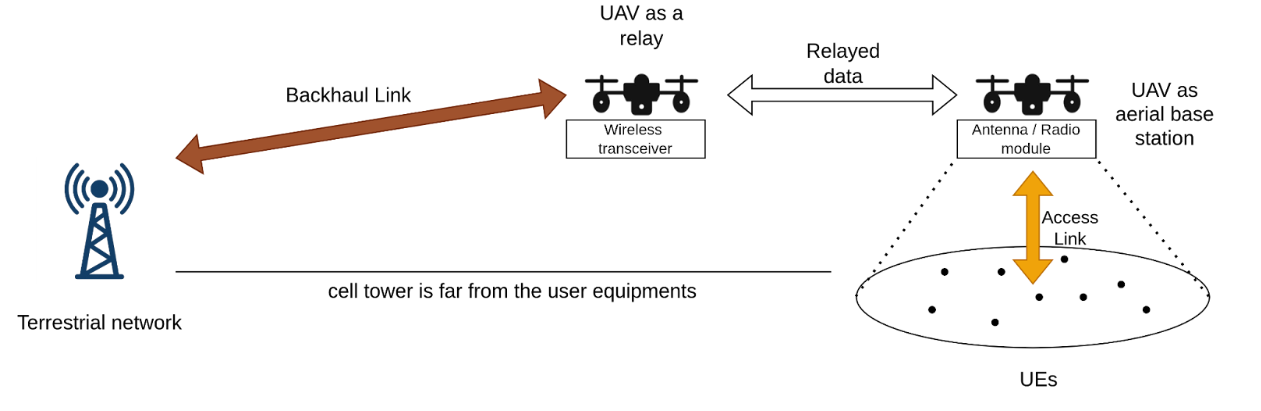}
    \caption{UAV as a relay node}
    \label{fig:uavr}
\end{figure*}

Relays in communication networks act as intermediate nodes, facilitating data transmission between the source and the destination. Relay nodes are especially helpful when there is a hindrance in the direct communication link between the source (origin of the transmitted signal) and the destination (intended recipient). Hindrances can be caused by obstacles, degradation of signals. Adding these intermediate nodes can improve coverage, increasing the reliability, efficiency, and quality of the signal. Some popular relaying schemes are Amplify and Forward (AF), Decode and Forward (DF), Compress and Forward (CF), selection relaying, demodulate and forward, and single-hop and multi-hop. More information about them can be found in  Tables \ref{tab:6} and \ref{tab:7} \cite{ref33,ref34,r1,r2,r3,r4}. Figure \ref{fig:uavr} is an example of a dual hop relay system. The terrestrial network (source) provides a backhaul connection to the first UAV, which acts as a relay bridge to cover the distance between the terrestrial network and the UAV acting as an aerial base station. After data is received from the backhaul relay, it forwards it to the aerial base station, which using an antenna/ radio module (payload) creates an access link for the users. 

\begin{table*}[htbp]
\centering
\caption{Types of relays}
\renewcommand{\arraystretch}{1.3}
\label{tab:6}
\begin{tabular}{|p{3cm}|p{2.3cm}|p{2.3cm}|p{2.5cm}|p{2.3cm}|p{2.5cm}|}
\hline
\textbf{Feature}
& \textbf{Amplify-and-Forward (AF)}
& \textbf{Decode-and-Forward (DF)}
& \textbf{Selection (Selective) Relaying}
& \textbf{Compress-and-Forward (CF)}
& \textbf{Demodulate and Forward} \\
\hline

Definition
& Relay amplifies the received signal that includes noise present in the system and forwards it towards destinations.
& Received signal is  filtered for noise, re-encoded and forwarded towards destinations.
& Relay forwards the received signal only after the decoding conditions are met
& Relay compresses the received signal and forwards compressed signal
& Demodulates the received signal and forwards the demodulated towards destination \\
\hline

Processing Complexity
& Low (Amplification)
& High (decoding + filter + re-encoding).
& Medium (signal evaluation+ relay selection).
& Medium to high (compression and forwarding).
& Lower than decode and forward \\
\hline

Noise Handling
& Amplifies signal and noise.
& Removes noise.
& Depends on relay selection and channel conditions.
& Noise not decoded; mitigated at destination via joint decoding.
& Considers demodulation errors based on finite parameters such as (SNR, fading, CSI) \\
\hline

Latency
& Low (immediate forward).
& Higher (decode/re-encode delay).
& Varies
& Moderate.
& Less than Decode and Forward \\
\hline

Power Efficiency
& Energy-efficient.
& Less efficient (decoding requires power).
& Adaptive.
& Variable
& Higher than DF \\
\hline

Reliability
& Lower 
& High due to signal regeneration.
& High.
& High
& Lower than DF \\
\hline

Error Propagation
& High.
& Low.
& Less 
& Low.
& Less than DF \\
\hline

Flexibility
& Simple scenarios
& Needed in highly efficient systems
& Highly flexible
& Suitable for dynamic/fading environments.
& Can accommodate different levels of CSI. \\
\hline

\end{tabular}
\end{table*}

\begin{table*}[htbp]
\centering
\caption{Single-Hop vs Multi-Hop}
\renewcommand{\arraystretch}{1.3}
\label{tab:7}
\begin{tabular}{|p{4cm}|p{5cm}|p{5cm}|}
\hline
\textbf{Features}
& \textbf{Single-Hop}
& \textbf{Multi-Hop} \\
\hline

Definition
& Direct transmits from source to destination
& Transmission happens via K intermediate nodes \\
\hline

Processing Complexity
& Less as only 2 nodes are involved
& High as multiple nodes are involved \\
\hline

Noise Handling
& Limited as only 2 nodes are involved  
(source and destination)
& Short distances between nodes can help in containing the noise \\
\hline

Latency
& Low
& Higher with each hop \\
\hline

Power Efficiency
& High for shorter ranges
& Low \\
\hline

Reliability
& Depends on the link between source and destination
& Performs better at long ranges if throughput is flexible \\
\hline

Error Propagation
& Direct link so minimal 
& High as with loss of one hop, packet is lost \\
\hline

Flexibility
& Limited range
& Flexible \\
\hline

\end{tabular}
\end{table*}

UAVs are emerging as a more flexible and reliable option for relaying in networks due to their ability to dynamically position themselves to improve network coverage, reliability, and QoS \cite{ref35,ref36}. UAVs acting as relay nodes typically carry a payload that includes transceivers, antennas to transmit and receive signals, and an on-board signal processing module. UAV based relays offer several advantages as compared to traditional networks, such as acting as a critical node in non-traditional network setups where deployment of fixed network infrastructure is not possible due to disaster or terrain challenges. UAVs, due to their ability to hover and maintain an elevated position, increase the probability of LoS for air-to-ground (A2G) and air-to-air (A2A) communications, which reduces the path loss and propagation challenges like shadowing \cite{ref36,ref37,ref38}. Despite these advantages, UAV based relays face various challenges that include limitations in endurance due to constraints on Maximum Take-Off Mass (MTOM), which restricts power source and payload carrying capacity, trajectory planning limitations, varying conditions of A2G channel that are influenced by Doppler effect, mobility, and environmental conditions (weather, terrain dynamics) \cite{ref39,ref40}. These constraints directly impact relay design choices, influencing the suitability of AF versus DF strategies and necessitating careful trade-offs between performance, energy efficiency, and computational complexity in UAV-assisted relaying networks.

The authors of \cite{ref41} discuss UAV as an AF relay for terrestrial networks based on the Friis transmission equation given in (\ref{eq:1}):
\begin{equation}
\begin{aligned}
P_r\,[\mathrm{dB}] = P_t\,[\mathrm{dB}] + G_t\,[\mathrm{dB}] + G_r\,[\mathrm{dB}] \\
- 20 \log_{10}(d) - 20 \log_{10}(f)
- 20 \log_{10}\!\left(\frac{4\pi}{c}\right)
\end{aligned}
\label{eq:1}
\end{equation}
where:
\begin{itemize}
  \item $P_r$ [dB] is the received power at the receiver;
  \item $P_t$ [dB] is the transmitted power from the source;
  \item $G_t$ [dB] is the transmitting antenna gain;
  \item $G_r$ [dB] is the receiving antenna gain;
  \item $d$ is the distance between the transmitter and receiver (in meters);
  \item $f$ is the carrier frequency of the transmitted signal (in Hz);
  \item $c$ is the speed of light in vacuum ($3 \times 10^8$ m/s).
\end{itemize}

\begin{figure*}
    \centering
    \includegraphics[width=0.75\linewidth]{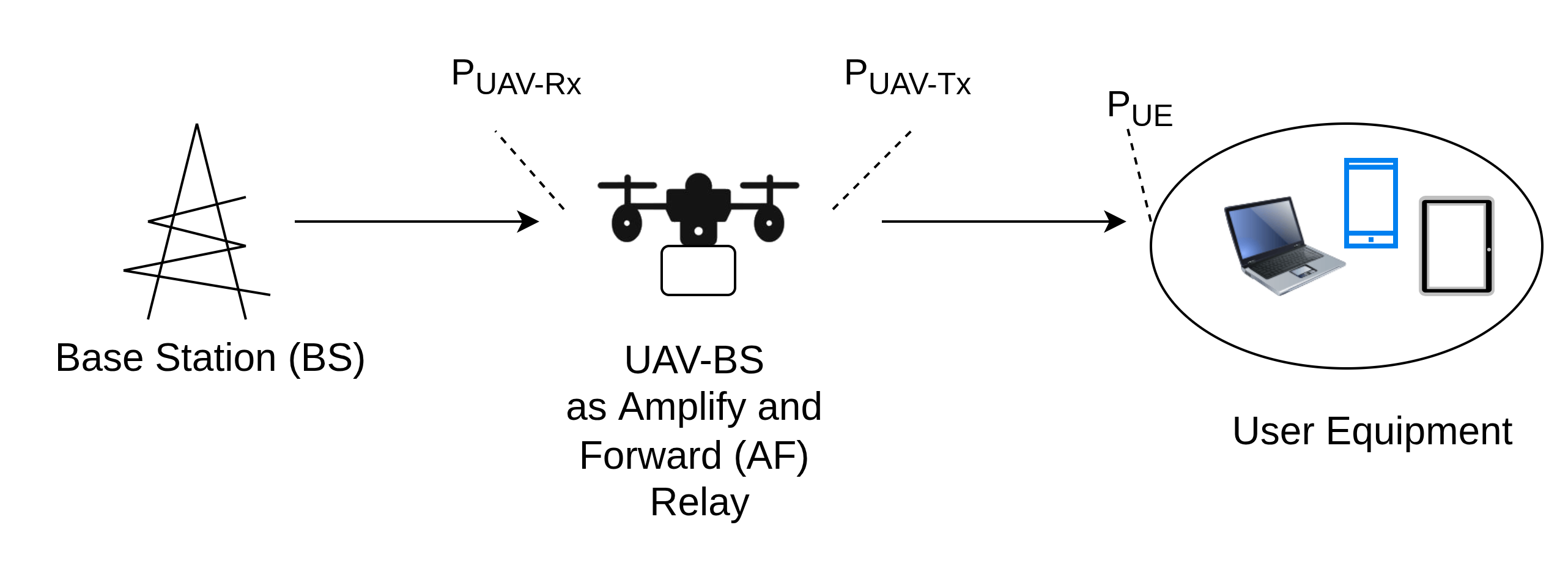}
    \caption{BS-UAV-UE end to end representation}
    \label{fig:6}
\end{figure*}

Data transmission as illustrated in Figure \ref{fig:6}, is a single hop communication link (BS-UAV-UE) where the UAV-mounted base station (UAV-BS) operates as an intermediate relay between the ground base station (BS) and user equipment (UE). The ground BS transmits the downlink signal to the UAV, which receives the signal with power. The UAV then amplifies the received signal and forwards it to the UE with amplified transmit power. The following assumptions are made:
 \begin{enumerate}
     \item The UAV hovers at a fixed position and operates as an AF relay, without any additional processing (digital operations more than amplification like modulation/ demodulation and beam alignment).
     \item The BS-UAV (backhaul link) and UAV-UE (access link) operate in mmWave band.
 \end{enumerate}
Based on these assumptions, the power received at the UE from the BS track is expressed as in (\ref{eq:2}):
\begin{equation}
\begin{aligned}
P_{\mathrm{UE}} = P_{\mathrm{UAV\text{-}T}} + G_{\mathrm{UE}} + G_{\mathrm{UAV}} + G_{\mathrm{BS}} \\
+ 20 \log_{10}\!\left(\frac{\lambda_1}{4\pi D_1}\right),
\end{aligned}
\label{eq:2}
\end{equation}
where:
\begin{itemize}
  \item $P_{\mathrm{UE}}$ is the received power at the UE;
  \item $P_{\mathrm{UAV\text{-}T}}$ is the transmit power of the UAV;
  \item $G_{\mathrm{UE}}$ is the antenna gain of the UE;
  \item $G_{\mathrm{UAV}}$ is the antenna gain of the UAV;
  \item $G_{\mathrm{BS}}$ is the antenna gain of the base station;
  \item $\lambda_1$ is the wavelength of the transmitted signal;
  \item $D_1$ is the distance between the UAV and the UE.
\end{itemize}
The authors of \cite{ref41} carried out simulations and demonstrated that the UAV altitude and horizontal positioning affect the received signal power at the UE. UAVs operating at higher altitudes (50 m and above) experience an increase in path loss and weakened received signals. A large horizontal separation between the base station and UAV reduces the received power at the UE. Simulations with elevation of approximately 30 m and horizontal distance range of 20-60 m provide a favorable balance between probability of LoS and signal strength. Using UAV as AF relay enhances the received power at UE and helps to cover gaps in the network coverage. The operational challenges that need to be addressed such as energy consumption, regulations, security risks, and connectivity need further research. 

\begin{figure*}
    \centering
    \includegraphics[width=0.75\linewidth]{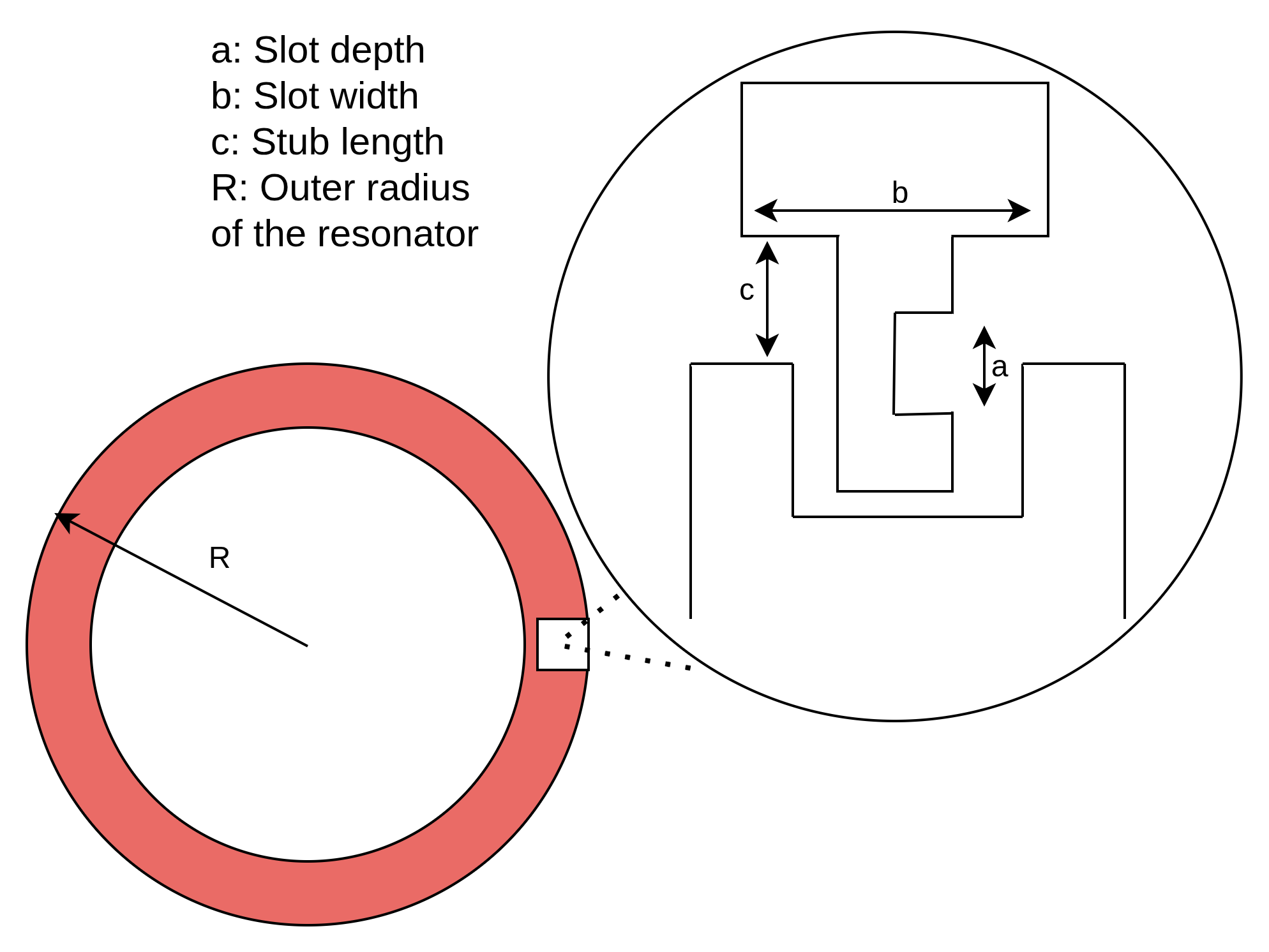}
    \caption{MTM Resonator}
    \label{fig:7}
\end{figure*}

To build further on the above insights, the authors in \cite{ref42} address the connectivity related issues identified in \cite{ref41} by emphasizing the optimization of key parameters like elevation angle,
environmental conditions (operational environment, propagation scenarios and weather) and UAV physical attributes (payload capacity, antenna configuration and placement and on-board power constraints). This enables UAVs to be deployed as effective relay nodes. Since obstacles in real time are not always deterministic, a probability-based LoS model is often put forth. Accurately modeled communication systems based on these factors achieve low latency, minimum signal loss, and high data rates. In addition to the environmental and geometric configurations for UAVs, the antenna design of a UAV also plays a vital role. A perfectly designed antenna can radiate/ beam-form precisely towards the destination, which can support a reliable LoS path and also enable techniques like MIMO, as proposed by the authors in \cite{ref43}. The authors propose an antenna design for 5G communications at 28 GHz, specifically for aerial base stations using Meta Materials (MTM), which are artificially designed materials to incorporate features like negative refractive index (RI) (effective permittivity and permeability is negative over a frequency range, which makes EM Waves refract in the opposite direction of that seen in traditional materials) and electromagnetic wave manipulation (control over the direction, phase, and behavior). Antennas used are $2 \times$ 8 MIMO antennas. The MTM resonator is designed using split ring resonator technology on a FR-4 Dielectric substrate (glass-fiber epoxy dielectric material widely used as an antenna), as can be seen in Figure \ref{fig:7}. The resonator is defined by its outer radius ``R", forming the main circular ring, while a zoomed-in view shows the detailed T-shaped slot geometry etched into the resonator. The slot is characterized by three key dimensions: slot depth ``a", slot width ``b", and stub length ``c". These parameters play a critical role in determining the resonator’s electromagnetic behavior, influencing its resonant frequency, impedance, and coupling properties. Based on the split-ring resonator design, it offers low reflection losses and also ensures low interference among the antenna elements. Their simulation results show that the design is very reliable and efficient. Thus, introducing meta-materials in antenna design can offer high gain, directivity, and enhancement in bandwidth that improves LoS while being lightweight and compact (suitable form factor for UAV deployment). A direction for future work is to build on the advantages of meta-material antennas with MIMO design to achieve higher data rates and better interference management, making them suitable for swarm UAV networks, integration with 6G, and self-sustaining base stations. Some areas such as increased path loss at mmWave frequencies, power management, integration of active and passive components, and signal degradation can be further improved \cite{ref43}.

Another application of UAVs as relays is the deployment of tethered UAVs in maritime emergency situations, in which the primary goal is to extend the network coverage that enables transmission of data in real-time. UAVs are connected to the vessel of the ship via a hybrid tether. A hybrid tether provides continuous power and high speed data (a tether eliminates the energy limitations of a UAV). The communication payload consists of mesh networking modules (a network topology in which each node can connect directly to more than one node, there is no single central access point, and nodes cooperate with each other and forward the data) featuring a 5 dB omnidirectional antenna with transmit power below 30 W and weight under 3 kg, 4G LTE modules for broadband voice and multimedia access compatible with mobile operators, and automatic identification System (AIS) receivers (GPS unit with data radio unit to identify ships and exchange important information among the ships) for real-time ship tracking and identification.  The UAV platform is a carbon-fiber hexacopter with six brushless DC (BLDC) motors capable of generating up to 8 kg thrust per propeller pair, controlled by an ARM processor integrated with gyroscopes, accelerometers, magnetometers, barometers, and Beidou GPS for precise navigation and flight stability. The tether system supports storm-grade 8 wind resistance and includes power conductors rated at 220 VAC and 30 kW capacity, alongside RS232/422 interfaces for telemetry and mission control. After a comparative analysis between traditional networks and UAV relay systems, it was observed that UAVs can increase the system communication range by approximately 50\%, which overall reduces signal attenuation \cite{ref44}. Although many benefits are offered by tethered UAVs, they are constrained by tether length, environmental conditions (high winds, moisture content in the atmosphere), payload constraints (tether weight, aerodynamic drag, and stability considerations). Improvements for these limitations is possible with the introduction of longer tether length and better tether management (drag reduction, active tension control, and stability).

Some other studies are compared in Table \ref{tab:r}.
\begin{table*}[htbp]
\centering
\small
\setlength{\tabcolsep}{4pt}
\caption{Comparative Analysis of UAV Relay and Multi-UAV Communication Studies}
\label{tab:r}
\begin{tabular}{|c|p{6.2cm}|p{3.6cm}|p{3.6cm}|}
\hline
\textbf{Ref.} & \textbf{Summary} & \textbf{Advantages} & \textbf{Limitations}\\
\hline
 \cite{rel1} & Addresses the security of relaying UAVs using Stackelberg game model where leader (UAV) first moves and follower (Jammer) follows the leader to achieve Stackelberg Equilibrium which in this case is to maintain the link between the users and UAVs without draining the battery, a higher secrecy rate is achieved & Improves signal reliability, shows higher efficiency than traditional methods & High computational overhead \\
\hline
 \cite{rel2} & Control the chain formation of multiple UAVs while bridging the gap between source and destination using a maintenance control algorithm (Consensus-Based Cooperative Control Algorithm based on two primary inputs- Relative UAV position and velocity matching) that ensures that each drone in the chain stays within the optimal distance of its neighbors to prevent signal dropouts & Extends the capabilities beyond single UAV & Chain is fragile, if single UAV in chain fails entire swarm collapses\\
\hline
\cite{rel3} & Discusses the physical design and testing of a multi rotor UAV designed for relaying. The hardware used for relaying is Wireless Relay Module that uses COFDM (Coded Orthogonal Frequency Division Multiplexing) wireless transceiver and ARM-based controller to monitor and manage data with omnidirectional antenna system to maintain link even if drone rotates & Stable hovering and reliable relay capabilities & Limited endurance \\
\hline
 \cite{rel4} & Explores use of UAV in tactical military networks, discusses about the various security protocols, encryptions, challenges while maintaining high broadcasting signal strengths in combat zones &  \begin{enumerate}
     \item Tactical coverage and BLoS connectivity- rapid restoration of communications, flexible repositions, reliable C2 link.
     \item A2G link- Increase in reliable link (LoS probability increases), path behavior is predictable, improved SINR.
     \item Swarm of UAV- fault tolerance, load sharing, healing networks. 
 \end{enumerate} & \begin{enumerate}
    \item Tactical coverage and BLoS connectivity – limited endurance, can be detected.
    \item A2G link – elevation-angle dependence, mobility induced variations.
    \item Swarm of UAV – inter-UAV interference, coordination overhead, multi UAV latency.
\end{enumerate}  \\
\hline
 \cite{rel5} & This paper studies the use of fixed wing UAVs as a relay between source and destination. It is studied for full duplex operations using AF relaying method due to requirement of being in constant motion without assumption of UAV flying in perfect vacuum and horizontal wind force acts on it and combining these with Self-Interference Cancellation (SIC) (removal of UAVs transmission) achieves higher Average Ergodic Capacity & Achieves much higher spectral efficiency, more realistic performance data by accounting for weather & Hardware susceptible to self interference \\
\hline
 \cite{rel6} & A hybrid link that uses Free Space Optics (FSO) (FSO near infrared optical band) laser based communication as the primary link, with Radio Frequency (RF) as a fallback and UAV acts as an intelligent relay which switches between these mechanisms based on link quality (instantaneous SNR, attenuation and pointing errors as lasers have narrow beans so UAV vibrations can disrupt these beams) & Massive bandwidth of lasers and the weather-resilient reliability of radio waves & FSO requires precise alignment   \\
\hline
\cite{rel7} & Considering swarm of UAVs this paper discusses a 3 layer hierarchical system- sensing layer (IoT nodes), Relaying layer (LALEs that gather data) and backhaul layer focusing on Joint Deployment and User Association where matching game theory is used to decide which UAV will pick up the data in order to avoid collisions & Offers extensive coverage & Difficult to model channels as reflections cause interference \\
\hline
\end{tabular}
\label{tab:UAV-relay-summary}
\end{table*}

UAVs deployed as relay nodes can provide a rapid and flexible solution to extend network coverage and improve connectivity in challenging environments. By optimizing UAV position, altitude, and elevation angle, coverage gaps can be filled in both terrestrial and emergency (maritime) scenarios. However, several challenges exist, such as limited flight endurance, optimal position, channel variance, propagation loss, interference, latency, scalability in multi-UAV deployments, regulatory constraints, and security risks.

\subsection{UAV as a Reconfigurable Intelligent Surface (RIS)}
RISs are an emerging technology for enhancing signal propagation in wireless communication networks. An RIS payload uses programmable meta surfaces that can manipulate electromagnetic waves. They can control the phase, amplitude, and polarization of transmitted waves and steer them to their propagation path; they are useful in applications like beamforming, coverage enhancement, and energy efficiency in communication networks \cite{ref45}. 

An RIS primarily operates in two modes:
1) Reflective RIS: Reflects the incident wave towards the destination. 
2) Transmissive RIS: Lets the incident waves pass through the surface towards the destination \cite{ref46}.

Other transfer modes include scattering and absorbing modes \cite{ref47,ref48}. A summary of these modes is provided in Table \ref{tab:8}.

\begin{table*}[t]
\centering
\caption{RIS Modes}
\label{tab:8}
\renewcommand{\arraystretch}{1.2}
\begin{tabular}{|p{3cm}|p{6.5cm}|p{6.5cm}|}
\hline
\textbf{Mode} & \textbf{Definition} & \textbf{Application in UAV Systems} \\ \hline

Reflective RIS & It reflects incoming EM waves with controlled phase shifts, acting like a programmable mirror. & Redirects gNB signals to UAVs in Non-Line-of-Sight environments, to enhance coverage and credibility on the link. \\ \hline

Transmissive RIS & EM waves can be passed through and also the phase and amplitude are modified. & UAVs receive signals through RIS-equipped building windows or barriers. \\ \hline

Hybrid RIS & Combination of reflection and transmission on a single surface. & Flexible deployment where UAVs may fly around or behind the RIS surface, adapting to coverage. \\ \hline

Absorptive RIS & Designed to absorb the incident EM energy, minimize reflections, and control interference. & Reduce signal leakage in UAV. \\ \hline

Scattering RIS & Initiates multiple dimension reflection to follow multi-path. & UAV-based MIMO channel improved, useful in high-mobility states. \\ \hline

Anomalous RIS & Wave propagation is altered using Snell’s Law. & Multi-user UAV support and Beam forming at sharp angles. \\ \hline

\end{tabular}
\end{table*}

Extending this approach, a UAV integrated with an RIS payload offers great flexibility for the optimization of  wireless signal parameters. As UAVs are airborne, their elevated positions can increase the LoS probability, decrease attenuation from obstacles, and enable 3-D positioning of RIS payloads. Thus, by acting as RIS platforms, they can extend coverage, increase signal quality, and reduce latency in scenarios where the deployment of traditional infrastructure becomes impractical. The scenarios discussed later shall highlight the potential advantages of this setup.

\begin{figure*}
    \centering
    \includegraphics[width=0.75\linewidth]{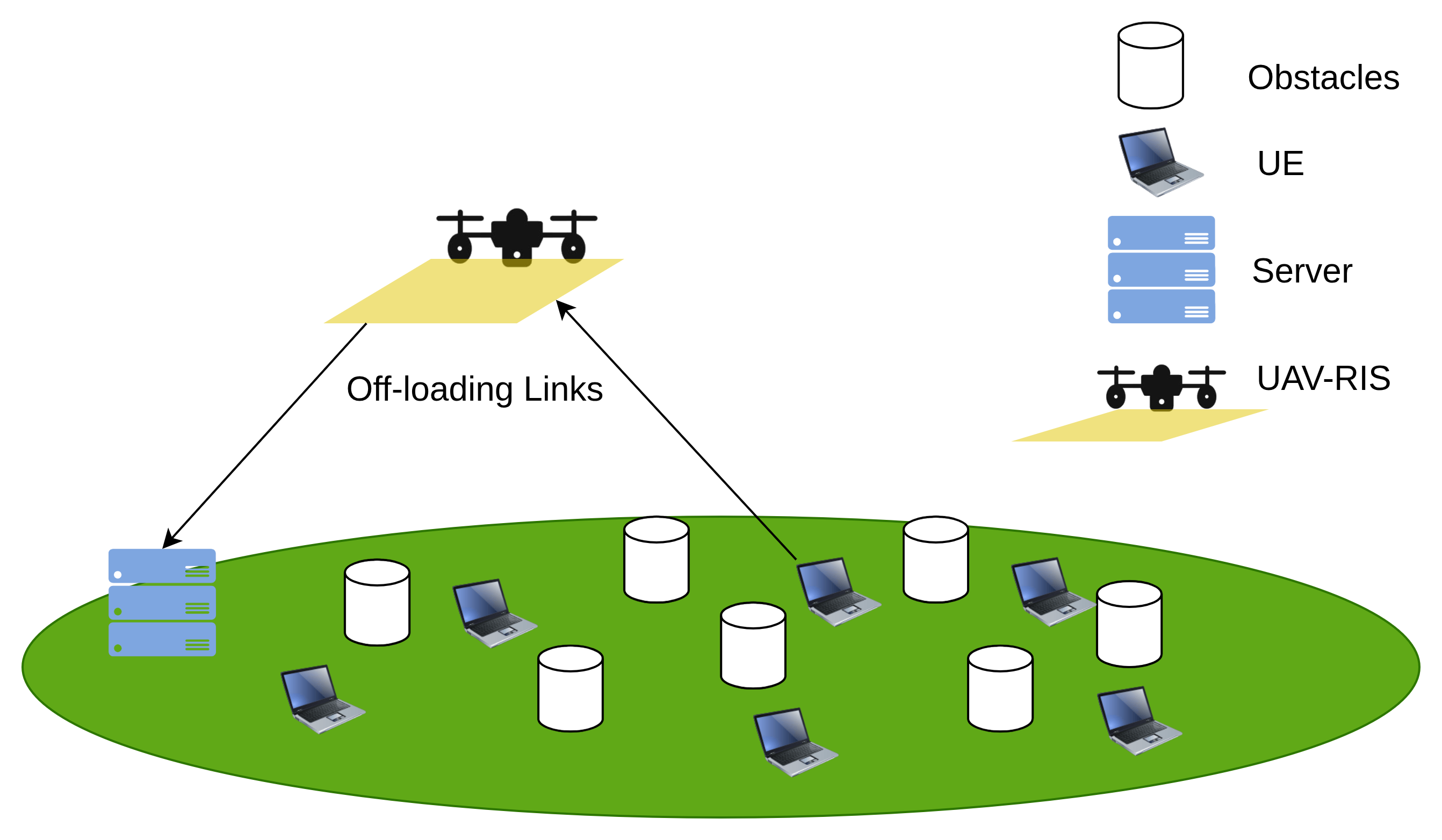}
    \caption{RIS-UAV System Model}
    \label{fig:8}
\end{figure*}

Mobile Edge Computing or Multi-Access Edge Computing (MEC) refers to the deployment of computation and storage resources at the network's edge, like base stations enabling low-latency processing and efficient task offloading for mobile users. MEC can be enhanced using a UAV assisted RIS setup. In this setup, a UAV carries a passive RIS payload. The payload’s orientation and position can be dynamically adjusted for signal optimization and can be propagated towards the MEC servers. The authors of \cite{ref49} propose an algorithm that maximizes the energy efficiency (EE) of the system (Energy efficiency is defined as the ratio of the total computed tasks (bits) to the weighted total power consumption of the system). This is done by enhancing the signal quality with the help of reflective surfaces that direct signals towards the Multi-access Edge Computing (MEC) servers. An RIS requires minimal energy to operate (passive), reducing power consumption significantly. This addresses the issue of high power consumption. The algorithm is illustrated in the system model shown in Figure \ref{fig:8}, in which there is a UAV-assisted communication framework with multiple UEs within a coverage area transmitting data to a remote server. The presence of obstacles may block the direct UE-to-server links, potentially degrading network performance. To mitigate this, UAVs equipped with RISs (UAV-RIS) are deployed above the coverage area, establishing offloading links that relay data from the UEs to the server. The UAV-RIS hovers at a height of $H$ (up to $100$ meters). The RIS has $N$ (up to $60$) passive elements and a MEC, which is responsible for offloading tasks to $K$ ground users, is at a fixed location. A Time Division Multiple Access (TDMA) model is used among users such that each user has an individual time slot. Similarly, the UAV motion is also divided into $T$ equal time slots. No direct user-server link is present; the energy consumption includes:
\begin{enumerate}
    \item Local computation and transmission energy (UEs); 
    \item Control Processing Unit (MEC);
    \item Propulsion energy for UAVs.
\end{enumerate}

Using Successive Convex Approximation (SCA), non-convex problems (optimization problems where the objective functions have multiple maxima/ minima) are approximated in a series of convex problems iteratively, and the solution is updated. Dinkelbach’s method is used to solve the rational programming method (iterative algorithm). The algorithm has three sub-modules: 
\begin{enumerate}
    \item UE scheduling and RIS phase shifts;
    \item UAV trajectory optimization and task allocation;
    \item CPU frequency selection. 
\end{enumerate}

Simulations are performed with 5 randomly placed users,  a UAV flying at 100 meters height, an RIS with 20-60 elements, and a MEC server located 400 meters away. Under these circumstances, their results show that UAVs hovering near high user density areas experience a higher energy efficiency. Employing more RIS elements can improve signal propagation. However, their algorithm has limitations as it assumes the following:
\begin{enumerate}
    \item Ground UEs are static;
    \item MEC server and other elements are ideal;
    \item Complete channel information is known;
    \item The links between UE-UAV with RIS and UAV with RIS-MEC server are quasi-static.
\end{enumerate}

Another application with UAV assisted RIS is the integration with a Simultaneous Wireless Information and Power Transfer (SWIPT) system that extends flight endurance. It considers a hierarchical time switching model, which provides separate phases for RF power harvest and data transmission. It employs a deep learning algorithm called Deep Deterministic Policy Gradient (DDPG), which adapts and learns continuous control strategies using an actor-critic architecture with target networks and previous replay experiences. The central RIS is used in data transmission, while other RIS elements are responsible for power harvesting. This approach is limited to a single UE \cite{ref50}.

\begin{figure*}
    \centering
    \includegraphics[width=0.75\linewidth]{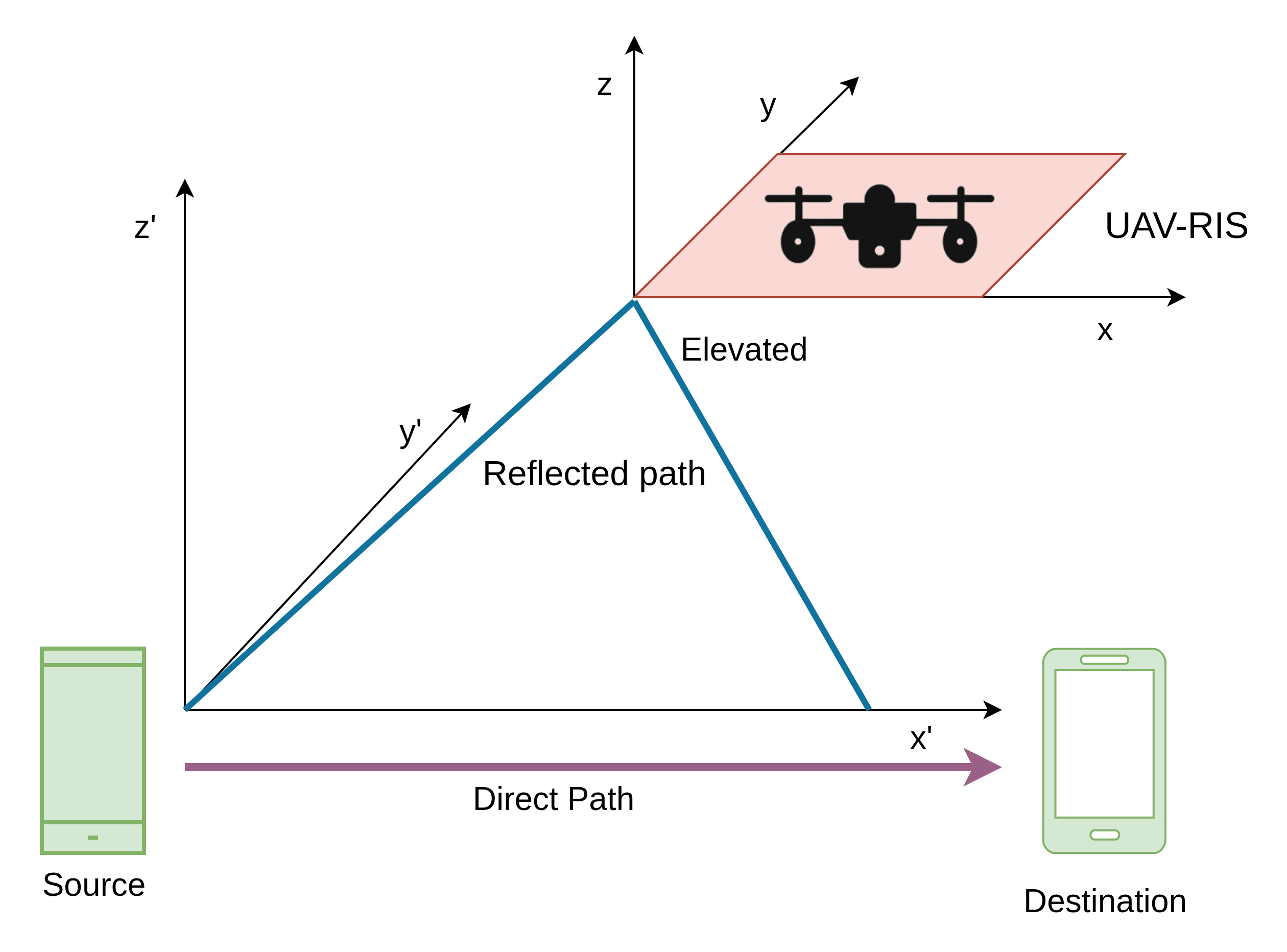}
    \caption{3 node system}
    \label{fig:9}
\end{figure*}

Trajectory optimization also plays a critical role in maximizing the communication performance of a UAV-RIS system. Here, a UAV’s position determines the quality of the hop for a UAV-RIS channel. By optimizing the UAV trajectory for phase-favorable locations and energy-efficient operating zones, the beamforming gains are enhanced, leading to improved SNR and overall system performance. The authors of \cite{ref51} consider an overall flight path using a 3 node communication model, as illustrated in Figure \ref{fig:9} that creates a geometric model of a UAV-RIS. It assists in wireless transmission between a source and a destination. In addition to a direct path, an elevated reflected path is established via the UAV-RIS system, which is positioned in three-dimensional space, defined by its $x$, $y$, and $z$ coordinates. The RIS reflects and reconfigures the incident signal from the source towards the destination, creating an alternative propagation path that can bypass blockages and improve link quality. The trajectory optimization is a non convex problem and is dependent on:
\begin{enumerate}
    \item Power optimization (UAV); 
    \item Reflection coefficient (RIS);
    \item Passive beamforming and trajectory optimization.
\end{enumerate}

For practical relevance, simulations are carried out with realistic constraints like minimum safety distance and feasible (maximum speed, allowed flying region, maximum acceleration) ways for a UAV to move from one place to another. The RIS is mounted on the UAV as a uniform planar array (UPA), where all the elements (RIS) are equally placed in horizontal and vertical directions in a grid layout. The phase shifts of these RIS elements are controlled by an RIS controller represented as a diagonal matrix that gets updated in each time slot. The channel between source and destination is considered as direct LoS and to capture more realistic effects (multipath), Rayleigh fading is applied.

Simulation results show that the structure yields higher throughput and better relaying performance by allowing the UAV to dynamically adjust its position to create more favorable channel conditions, rather than simply flying directly from the source to the destination or remaining in a fixed position. This improves link strength, SINR, and fairness of usage among users.

Although the above method improves communication performance, it puts significant computational load on the system, which overall increases the system complexity as multiple optimization parameters come into the picture. To address these limitations, another approach is considered that increases the UE throughput while also considering the limited power source availability. The authors of \cite{ref52} focus on optimizing critical parameters like UAV positioning, transmission power of the link, RIS phase shifting, and beamforming capacity of the system. The approach is as follows:
\begin{enumerate}
    \item RIS with $M$ Elements: A passive array that dynamically controls the wireless channel through phase adjustments; 
    \item UAV: Hybrid Access Point capable of downlink wireless power transfer (WPT) and uplink wireless information transfer (WIT); 
    \item UEs: Single-antenna devices with constrained battery life, relying on the UAV for both power and data transmission. 
\end{enumerate}

An SCA based optimization approach is considered. With only a few iterations, the entire area is covered by this framework, achieving the desired computational efficiency. The entire communication is divided into two time slots:
\begin{enumerate}
    \item WPT for harvesting energy (power supply);
    \item WIT for data transmission.
\end{enumerate}
The simulation results show that the proposed method is able to achieve the desired trajectory resource allocation performance in only a few iterations.

Multiple studies on UAV-RIS are ongoing as summarized in Table \ref{tab:ris}.
\begin{table*}[!htbp]
\centering
\small
\setlength{\tabcolsep}{4pt}
\caption{UAV as RIS}
\label{tab:ris}
\begin{tabular}{|c|p{8cm}|p{5cm}|p{4cm}|}
\hline
\textbf{Ref.} & \textbf{Summary} & \textbf{Advantages} & \textbf{Limitations}\\
\hline

\cite{ris1} & Discusses a Federated Deep Learning framework in which swarm of UAVs act as learners to predict channel conditions and mobility of users model that shares weight (Neural network weights/ gradients) updates instead of raw data. The system then optimizes RIS phases and UAV trajectory path to maximize the energy efficiency and data rate without compromising the privacy of ground-level users. & Keeps sensitive data on-device, Reduces communication overhead  & High on-board computational demand; requires frequent synchronization between the UAVs and the programmed model. \\ \hline

\cite{ris2} & Discusses a virtual partition strategy that divides a single RIS element into multiple independent surfaces to serve a swarm of UAVs simultaneously by creating reflection beams dedicated to each UAVs, using Block Coordinate Descent (iterative optimization method) and the Adam Optimizer (a gradient-based optimizer used in machine learning and deep learning) an optimal 3D placement of the RIS is found which facilitates the areas which need to be covered &  Supports swarms using a single infrastructure, eliminates blind spots in dense urban environments. & Reduced Array Gain \\ \hline

\cite{ris3} & Investigates the integration of RIS within (MEC) architectures. The model focuses on 3 technologies (RIS, MEC and UAV) and the effective ways to manage these technologies is by analyzing how the UAV’s flight path must be coordinated with the RIS’s reflection angles, splitting of power source energy in between propulsion and transmission and mitigation involved for interferences. & Extends mission time by outsourcing heavy computation; optimizes global system power consumption. & Primarily a theoretical survey\\ \hline

\cite{ris4} & Discusses the physical geometry, how the mechanical azimuth and elevation angles of a building-mounted RIS impact the signal throughput for a mobile UAV as on contrary most of the papers assume RIS to be a flat surface. Using SimRIS channel simulator model, the authors prove that physical tilting the RIS to track the UAV's trajectory provides a much higher "Signal-to-Noise Ratio" (SNR) than simply moving the UAV to a different location which is critical for high data transfers & Mechanical tilting can improve signal strength by 200\% compared to increasing transmit power. &Requires precision motors and gimbals which are susceptible to environmental wear and tear. \\ \hline

\cite{ris5} & Focused on disaster scenarios, it analyzes the use of UAV–RIS assisted communication links under stringent time constraints, ensuring successful transmission of a survivor’s phone GPS distress signal before connection timeout and link failure.  Authors develop a non-uniform codebook design for rapid beam alignment. With the help of Karush-Kuhn-Tucker (KKT)(set of necessary conditions for optimality in constrained optimization problems) conditions to solve the power minimization problem, the system allows a UAV and RIS to establish a secure communication link with a survivor's device in milliseconds, bypassing exhaustive search methods. & Establishes critical links in milliseconds; vital for time-sensitive search-and-rescue operations & The fast-track search can lead to sub-optimal beam pointing compared to exhaustive search methods. \\ \hline

\cite{ris6} & Focuses on Zero Energy IoT through the programmed model where UAV acts as an airborne reader for battery less sensors (smart agriculture field) RIS here, captures UAV radio signals like a magnifier and reflects them towards sensors, sensors then reflect this information back towards the UAV. The RIS significantly boosts these tiny reflections, which would otherwise be too weak for the UAV to hear.  & Enables a maintenance-free, green network ideal for remote oceanic or agricultural monitoring. & Signal decays exponentially, severely limiting the effective operational range. \\ \hline

\cite{ris7} & It addresses interference problems in Space Air Ground Integrated Network (SAGIN). When all the elements like satellite and the UAVs operate over the same frequency bands, it can lead to congestion and interference that degrades the quality of the network. Thus, authors use mathematical modeling interference alignment where a RIS mounted on an UAV cancels out the noise due to satellite by reflecting the satellite signal into a phase called null space which creates a clear path way to communicate Degrees of Freedom (DoF) is also analyzed to see how many simultaneous UEs can be handled by the system.& Allows satellite and 5G/6G to share frequencies without mutual data corruption. & Extreme difficulty in maintaining a 3-way alignment between a satellite, a flying UAV and an UE. \\ \hline

\cite{ris8} & Authors establish a framework for dual hop UAV-RIS communication systems for performance metrics such as Outage Probability (Probability of received SNR falling below the threshold) and Average Bit Error Rate (Probability that detection of the transmitted bit is incorrect due channel conditions like noise and fading) by statistically characterizing (different channel variations) the end to end SNR of channel under fading conditions. Simulation results validate the accuracy of the derived theoretical expressions. & Provides a mathematical Gold Standard for calculating optimal UAV altitude and RIS placement. & Does not fully account for physical UAV jitter or rapid atmospheric turbulence. \\ \hline

\end{tabular}
\end{table*}

UAV as a RIS platform can offer flexible and efficient solutions and enhance wireless communications by improving signal propagation, coverage, and energy efficiency. Using UAV maneuverability and RIS programmability, these systems can be used to optimize link conditions (received power, SNR, throughput). Practical challenges like computational complexity, accurate channel state information (CSI), scalability, energy efficiency, and robustness under dynamic environments where network conditions can change, remain significant considerations for practical deployment.

\subsection{UAV as a User Equipment (UE)}
\begin{figure*}
    \centering
    \includegraphics[width=0.75\linewidth]{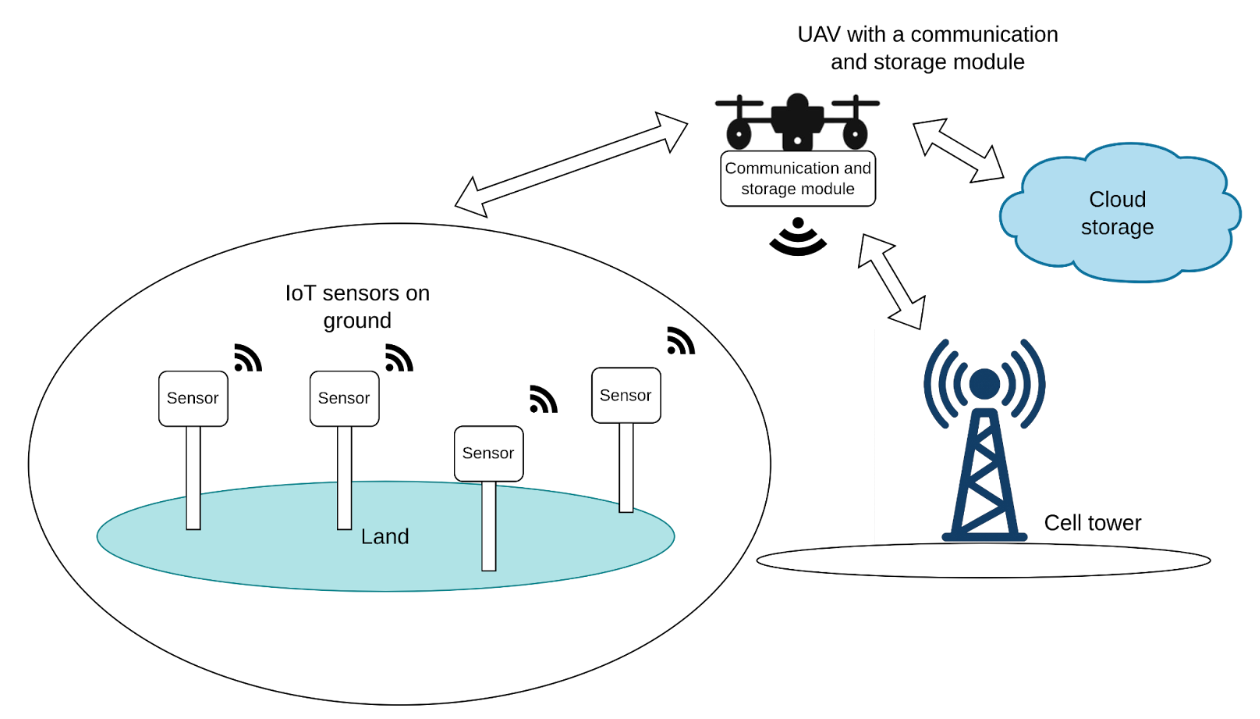}
    \caption{UAV-enabled IoT Data Collection and Storage}
    \label{fig:uavue}
\end{figure*}

A UAV can be integrated as a UE within a system that supports diverse functions such as surveillance, safety and monitoring, data acquisition, and wireless communications. When a UAV acts as a UE in a communication network, it carries an application compliant communication payload that may include an onboard modem, antennas aligned for aerial orientations, other supporting circuits (power amplifiers, RF switches, Filters, oscillators, etc.) and may also include a high resolution camera/ environmental sensors. UAVs engaged as UEs require a higher uplink throughput as they transmit sensor/ video data in real-time to the ground station while in motion. The example in  Figure \ref{fig:uavue} illustrates the UAV serving a gateway that collects data from multiple IoT sensors. Once collected, the UAV can offload this information to a cell tower or directly sync it with cloud storage for long-term processing and accessibility. This framework is more suited for monitoring a particular site.
This section highlights such scenarios in which a UAV operates as a UE within the communication system, enabling the operations discussed previously. 

An example of  a UAV being used for inspection is presented by the authors of \cite{ref53}. They investigate the structural health of a bridge using  a UAV instead of any manual inspection, where trained professionals are expected to access the structure, exposing them to safety risks with the possibility of human errors. The method consists of a mobile based acquisition that uses a shelled (collision resistant) UAV.
The UAV includes:
\begin{enumerate}
    \item A protective shell that is collision resistant (useful in confined space) and is waterproof;
    \item A gimbal for stable orientation control;
    \item Onboard camera for streaming of data (bridge video) in real-time;
    \item Communication interface that is linked to the user application (application developed by the Department of Public Works and Highways).
\end{enumerate}
The workflow of the proposed system is divided into six stages, which  are as follows:
\begin{enumerate}
    \item Pre-flight checks;
    \item Drone deployment and operation;
    \item Data Collection and storage;
    \item Damage detection;
    \item Inspector evaluations;
    \item Report creation.
\end{enumerate}
The UAV here is not a UE, but an extension of a UE, which helps with live streaming and interaction with the inspection application developed by the authors of \cite{ref53}.

Another application, proposed by the authors of \cite{ref54}, involves equipping a UAV with  a front of view (FoV) camera that enables real-time Virtual Reality (VR) streaming of the aimed area over 5G. This method leverages both Enhanced Mobile Broadband (eMBB) and URLLC of 5G. Using Content Delivery Networks (CDNs) and Mobile Edge Computing (MEC), the end-to-end latency (drone-user’s VR headset and vice-versa) was found to be less than 20 ns. This configuration enables remote inspection in real-time. The inspection is carried out using the Popularity Based (PB) method, a routing strategy where the desired area is divided into multiple geographic layers, directing messages towards the area layer that is closest to the destination. However, this approach uses a single UAV and the system is constrained  due to limitations like limited power supply and reduced ability to cover larger areas. 

To overcome these limitations, multi-UAV systems could be considered, e.g., the authors in \cite{ref55} propose a system that is designed for agricultural sensing tasks. A swarm of quadcopter UAVs with Robot Operating System (ROS), open source middleware frame that is used for testing, developing, and running robot applications is considered. Each copter in the system is incorporated with  Red, Green, Blue (RGB) cameras,  Inertial Measurement Units (IMUs), and  Global Positioning System (GPS) for sensing. The systems use a velocity-based control law to have an individual UAV control (autonomous or tele-operated) formation control to maintain safe distances and coordinated behavior, and obstacle avoidance through fields. A wireless network is used to establish communication. The study is conducted in four formations with three trials in each (12 trials). The configurations used are: 
\begin{enumerate}
    \item Auto-Single-UAV (autonomous single UAV);
    \item Auto-Multi-UAV (autonomous swarm of UAVs);
    \item Tele-Single-UAV (manually operated single UAV);
    \item Tele-Multi-UAV (manually operated swarm using haptic feedback and swarm control).
\end{enumerate}
The simulation results show that the task completion time is reduced by 58.7\% for  a swarm of UAVs (tele-operated), while automated operations showed  that the task completion time is reduced by 80\%. Other observations show that  tele-operated systems reduce the preparation and setup time significantly due to the absence of programming, whereas autonomous systems benefit from trajectory planning and efficient resource utilization. Overall, the performance metrics play a vital role in determining mission and operational objectives. More such studies are included in Table \ref{tab:ue}.

\begin{table*}[!htbp]
\centering
\small
\setlength{\tabcolsep}{4pt}
\caption{UAV as UE}
\label{tab:ue}
\begin{tabular}{|c|p{8cm}|p{5cm}|p{4cm}|}
\hline
\textbf{Ref.} & \textbf{Summary} & \textbf{Advantages} & \textbf{Limitations}\\
\hline
\cite{ue1} & This paper focuses on using UAV as UE to inspect and assess any damage to onshore and offshore wind power stations. UAVs are equipped with sensing platforms that have higher resolutions to capture structural and surface condition data of the wind turbine blades and towers. To carry out these types of studies the focus is on flight planning, data collection and post processing of the data. Simulations involved use of multi rotor UAVs. The entire system is evaluated on factors like inspection coverage area, time efficiency, and detection capability. & Significant increase in operational safety by removing human climbers; provides cost-effective monitoring for remote offshore locations. & High sensitivity to wind gusts and potential electronic failure due to salt-spray corrosion \\ \hline

\cite{ue2} & This paper uses UAV to detect tree hazards near power lines using multi-source remote sensing, The UAV is equipped with Li-DAR and RGB cameras to acquire data. Machine learning or computer vision algorithms are applied to detect vegetation encroachment and to assess risk to power infrastructure. & Enables large-scale automation of power line maintenance; significantly reduces the risk of wildfires and outages. & High demand for computational power to process fused Li-DAR and multi spectral datasets. \\ \hline

\cite{ue3} & Authors proposes an authentication scheme for UAV assisted UE, where UAVs act as aerial network nodes to support authentication and access procedures. UAV act as trusted intermediary and authenticates itself to the core network. Once authenticated, nearby UEs are authenticated by UAV which relays their (UEs) credentials to the core network. The scheme ensures mutual authentication, identity privacy, and resistance to quantum and traceability attacks while keeping computation and signaling overhead low & Enhances network reachability in undeserved areas; maintains 6G-level security protocols for remote authentication. & Introduces a single point of failure if the proxy UAV is disconnected or runs out of battery. \\ \hline

\cite{ue4} & These studies focus on how cellular mobile networks can support large scale UAV services. The study focuses on challenges like scalability, interference, mobility and QoS. UAVs are assumed to be equipped with LTE/5G user equipment modules and directional or omnidirectional antennas. Simulations are carried out to analyze factors like throughput, latency, and network capacity. Results show that network can support large UAVs to a certain point. Services requiring strict QoS, mobility, and integration with traffic management systems need further evolution in both mobile network capabilities, standards, and integration architectures. & Facilitates seamless integration of commercial UAVs into cellular grids; improves resource management for aerial services. & Frequent uplink interference caused by UAVs connecting to multiple base stations simultaneously. \\ \hline

\cite{ue5} & This paper discusses a method that help in detection of airborne UEs. To detect these UEs it uses factors such as signal strength, handover behaviors and timing pattern that differs ground from aerial UEs. All the UEs are assumed to be connected to LTE network. The method is based on network management (identification of UEs and manage their interference, mobility, and resource allocation), simulation and field results from the LTE equipments. & Provides regulatory oversight for mobile operators; requires no hardware modifications to existing LTE towers. & Potential for false positives in high-altitude terrain or mountainous regions. \\ \hline

\cite{ue6} & This paper uses ML integrated with UAV for content creation. The UAV is equipped with cameras and onboard ML processing module in order to enhance capturing of video, content analysis. This method targets production of multimedia data. Experimental usage of ML with UAVs is seen to give higher content quality and improved scenario handling.  & Revolutionizes content production through autonomous framing; reduces the need for manual piloting in media sets. & Intense computational overhead on-board reduces flight times of the drone. \\ \hline

\cite{ue7} & This paper focuses on transition of UAV between water and air. he paper analyzes dynamic behavior during water exit and aerial transition. Validation is typically performed via physical experiments and/or simulations of vehicle dynamics & Enables multi-domain missions (underwater and aerial); provides a platform for clandestine marine research. & Immense mechanical stress during the transition phase often causes structural fatigue or water ingress. \\ \hline

\cite{ue8} & This survey paper reviews challenges in autonomous UAV cinematography. It discusses UAV platforms equipped with cameras, gimbals, and onboard perception and planning systems for automated filming. The paper does not present a single hardware platform but surveys existing systems and prototypes. Simulation and experimental platforms referenced in the literature include quadrotor drones with vision-based tracking, onboard GPUs or embedded processors, and trajectory planning algorithms. The UAVs are primarily sensing and filming platforms for content creation (multimedia) , with communication used for control, monitoring, and video streaming rather than as communication relays or base stations. & Enables complex aerial shots that are impossible for human pilots; improves on-set safety against obstacles (infrastructure, people) through advanced avoidance algorithm. & Significant processing bottlenecks when handling simultaneous 4K video encoding and autonomous navigation. \\ \hline

\end{tabular}
\end{table*}

Thus, UAVs can act as multiple communication nodes, and each role requires a set of parameters that can let the UAV work as that node. Table \ref{tab:9} summarizes some parameters for the UAV as a communication node, explained above.

\begin{table*}[htbp]
\centering
\caption{UAV Roles in Wireless Communication Networks}
\label{tab:9}
\renewcommand{\arraystretch}{1.2}
\begin{tabular}{|p{3.5cm}|p{2.8cm}|p{2.8cm}|p{3.2cm}|p{3cm}|}
\hline
\textbf{Parameter} & \textbf{UAV as gNB (Ground NodeB)} & \textbf{UAV as Relay} & \textbf{UAV as an UE (User Equipment)} & \textbf{UAV as RIS (Reconfigurable Intelligent Surface)} \\ \hline

Computational Load & Very High & Medium & Low–Medium & Very Low \\ \hline
Deployment & Static and hovering & Requires LoS & Mobility defined & Stationary \\ \hline
Antenna Type & MIMO & Directional & Directional & Passive (Metasurfaces) \\ \hline
Energy Consumption & High & Medium & Low–Medium & Low \\ \hline
Mobility Requirements & Mostly stationary & Alignment required & High portability & Low \\ \hline
Altitude Range & 100–300 m & 100–200 m & 50–150 m & Depends on coverage gap location \\ \hline
Latency & Highly sensitive & Medium & Depends on the tasks & Low \\ \hline
Complexity & Very High & Medium & Low & Medium \\ \hline
Use Case & Emergency gNB, rural extension & NLoS link bridging, relay & Aerial Surveillance & Overcome blockage \\ \hline

\end{tabular}
\end{table*}

\section{UAV power sources and their effects on communication payload}
UAVs in communications play a significant role in extending network coverage and enhancing network reliability as compared to ground communications infrastructure. However, optimization of UAV architectures to maintain the existing rate of efficiency of energy (power source), throughput, and coverage is desirable. Additionally, other factors like power management and power sources also play an important role in determining the flight endurance, payload capacity, and overall reliability in communication networks. This section discusses different types of power sources and their effects on various communication payloads.
\subsection{Brief Overview of Power Sources}
UAVs for communication related applications require efficient power sources to sustain flight operations and operate onboard communication payloads. The mainstream power source for UAVs is a Lithium Polymer (Li-Po) battery. Li-Po batteries are considered to be portable/lightweight and energy-dense in nature. Even though they are a reliable option, flight times using them are capped to a maximum duration of 90 minutes \cite{ref56}. Other lithium-based batteries such as Li-ion and LiFePO₄ are also used in UAVs. LiFePO₄ batteries offer a high energy density in contrast to Li-ion batteries. They provide thermal stability, longer battery life, and better safety due to a more stable crystal structure. Recent developments in carbon-coated LiFePO₄ cathodes further improve conductivity and high-rate performance \cite{bat3}. Tethered UAVs have evolved to be an alternative to Li-Po batteries, where a continuous power supply is ensured through specialized cables from ground sources like generators, grid power (AC), battery packs, and vehicle-mounted power packs. Hybrid tethered cables are also available, which can be used as  a power source and data transmission cables like optical fiber that enable uninterrupted communication between UAV and ground infrastructure \cite{ref57,ref58}. Fuel cells are also capable of providing longer flight endurance, but are limited due to their low energy conversion efficiency and requirement of the integration of a complex system of hydrogen storage \cite{ref59}. Internal combustion engines (ICEs) and renewable energy sources such as solar power are also being explored as alternative UAV power sources. Renewable energy sources enable long-endurance operations,  which are sustainable and environment friendly. Although ICEs can provide high energy density for large and long-range UAVs, they increase the noise and mechanical complexity of the entire system \cite{bat4}. A comparative study can be seen in Table \ref{tab:10}. Overall, the choice of power source is always influenced by flight endurance, desired capacity to carry payload, and system complexity (autonomous operations, computational load, etc.).
\begin{table*}[htbp]
\centering
\caption{Power sources for UAVs}
\label{tab:10}
\renewcommand{\arraystretch}{1.3}
\resizebox{\textwidth}{!}{%
\begin{tabular}{|p{3.2cm}|p{5cm}|p{5cm}|p{5cm}|p{6cm}|}
\hline
\textbf{Power Source}
& \textbf{Energy / Power Characteristics}
& \textbf{Advantages}
& \textbf{Limitations}
& \textbf{Implications for UAV Communication Roles} \\
\hline

Li-Ion / Li-Po Batteries
& Moderate energy density (150–260 Wh/kg); offer limited endurance
& Well-developed technology, lightweight, simple integration
& Short flight time, slow response to peak loads
& Limits relay/gNB operation and high RF transmit power \\
\hline

Super capacitors (with batteries)
& Very high power density; low energy density
& Excellent for high power and rapid changes in voltage and current in the circuit
& Cannot sustain long missions
& Supports high-power RF transmission and easy maneuvers in comm UAVs \\
\hline

Fuel Cells 
& Very high energy density
& Extended endurance, quick refuel, low noise
& Storing is complex, weight and cost
& Offers Continuous services (like relaying) \\
\hline

Hybrid Power Systems
& Combination of battery + fuel cell + supercapacitor
& Balances endurance and peak power; high reliability
& Increased system complexity, control overhead
& Best suited for mixed roles requiring both long endurance and high RF power \\
\hline

Renewable Energy Sources
& Energy harvesting with low instantaneous power; environment dependent
& Sustainable operation, extended endurance
& Low power density, weather and environment dependent
& Suitable for very long endurance, low-rate communication platforms \\
\hline

Tethered Power Supply
& Continuous power supply
& Infinite endurance, high payload support
& Limited mobility due to tether
& Ideal for static aerial operations and UAVs equipped with heavy payloads \\
\hline

Energy Management System (EMS)
& Multi-source power dispatch
& Optimized energy efficiency and flight endurance
& Control of power source complexity
& Critical for sustaining communication QoS and uptime \\
\hline

\end{tabular}%
}
\end{table*}

\subsection{Different Types of Communication Payloads}
\label{sec:B}
UAVs, which are part of a specialized communication operation, require different onboard payloads that are as follows:
\begin{enumerate}
    \item Software Defined Radios (SDRs) replace all traditional radio hardware with software. They are versatile, as they allow for real-time changes, such as frequency and spectrum management, which require only a software update; 
    \item Connectivity modules such as 4G/5G or beyond 5G modules, Wi-Fi chipset, IoT gateways/ sensors. These modules can help UAVs act as aerial access points for environmental surveillance and data acquisition; 
    \item Antennas to perform directional transmission of  signal, beamforming to offer throughput and LoS stability from UAV to terrestrial UEs;
    \item Satellite communication terminals for Beyond LoS (BLOS) communications \cite{ref60,ref61,ref62,ref63}. 
\end{enumerate}
The different types of communication payloads are summarized in Table \ref{tab:11}.
\begin{table*}[htbp]
\centering
\caption{Communication Payload Characteristics and Power Source Suitability}
\label{tab:11}
\renewcommand{\arraystretch}{1.2}
\begin{tabular}{|p{3cm}|p{2.5cm}|p{2.5cm}|p{2.5cm}|p{2.5cm}|p{2.5cm}|}
\hline
\textbf{Payload Type} & \textbf{Power Consumption} & \textbf{Weight} & \textbf{Thermal Management} & \textbf{Integration within the network system} & \textbf{Power Source} \\ \hline

 Software Defined Radios (SDRs) & 5–15 W & 200–400 g & Active Cooling & Medium & Hybrid System \\ \hline

Connectivity (4G/5G/Wi-Fi) Modules & 1–3 W & 150–250 g & Passive Cooling & Low & Li-Po Battery \\ \hline

Antennas & 5–10 W & – & Active Cooling & High & Tethered System \\ \hline

Satellite Communications & 10–25 W & 500 g and above & Heat Sink & Extreme & Fuel Cell \\ \hline

\end{tabular}
\end{table*}

\subsection{Overview of Integration of Power Sources and Communication Payload}
\begin{figure*}
    \centering
    \includegraphics[width=0.75\linewidth]{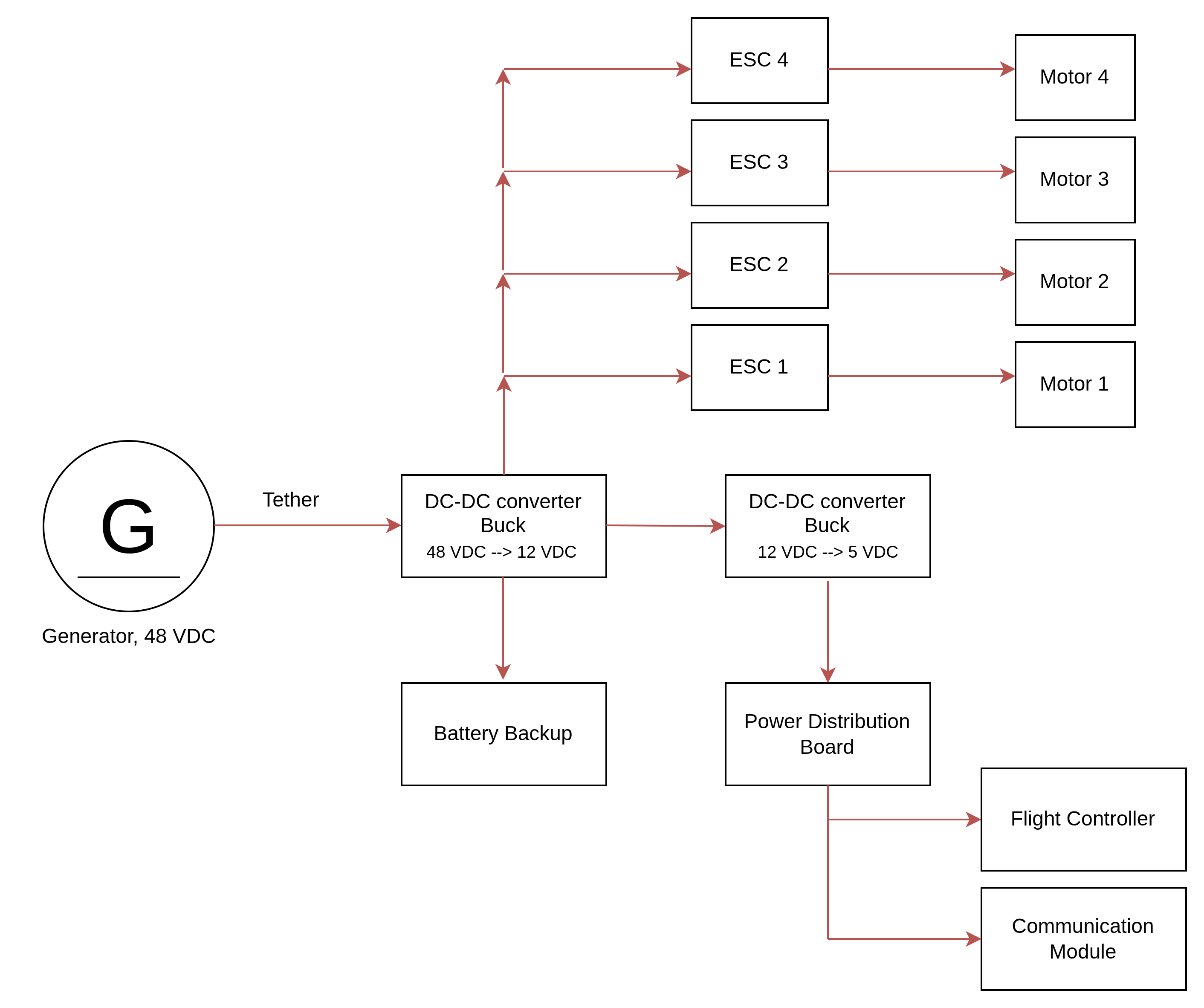}
    \caption{Example illustration of integration of power source with communication payload}
    \label{fig:11}
\end{figure*}

Figure \ref{fig:11} is an example illustration of the integration of a power source with a communication payload. A tethered power source is considered, keeping in mind the requirement of high flight endurance. Typically, for this type of integration \cite{ref55}, the power source consists of a generator that can be AC or DC. Generally, DC generators are preferred due to simple power conversion, stable transmission, and since  they do not need additional filters (rectification is required for AC to DC, LC/ RC filters for eliminating ripples, harmonics) followed by an onboard DC/DC converter, which is a buck converter that converts 48 VDC to 12 VDC for BLDC motors (electronic speed controllers); these motors are responsible for propulsion and battery backup. Another buck DC/DC converter is deployed, which converts 12 VDC to 5 VDC that is routed through the Power Distribution Board (PDB) to the Flight Controller (FC) and communication module; all these modules are interchangeable with the payload depending on the functionality/ role in a task \cite{ref64,ref65,ref66}.

\subsection{Power Source Suitability for Communication Payloads}
Communication payloads, discussed in Section \ref{sec:B}, have different power requirements and are suited to different power sources, as the energy from the power source is a combination of power supplied to the propulsion system, power supplied to the avionics, and power supplied to the payload (unless the payload has its own power supply system). Hence, for communication payloads like repeaters/ short-range, which are lightweight and less power-intensive, Li-Po batteries are the best fit. Meanwhile, SDRs and relays can be equipped with tethered/ hybrid systems. Fuel Cells are suitable for applications requiring endurance for long-range payloads; but due to the complexity of their hydrogen storage system, they are limited in applications. Table \ref{tab:11} summarizes the communication payloads and power source suitability for various UAVs \cite{ref67,ref68,ref69,ref70}.

\section{Security in UAV Communication}
Recent research focuses on UAV communications and their widespread use in domains such as military, surveillance, and public safety, where secure and reliable communication is necessary. UAVs face several communication threats such as eavesdropping, interference, spoofing, Man in the Middle Attack (MitM) and control hijacking. Many commercial UAVs still lack encryption and identity authentication (authorized UAVs like pre-registered identifiers like UAV operator registration number only have access) and when interacting with other types of communication nodes (IoT, cloud), attackers have a chance to exploit weak spots in the communication links present between the UAVs and other nodes in the network. Data transmission should satisfy core attributes such as confidentiality, integrity, authenticity, and availability (for critical communication links). UAV data security can be classified into various sensitivity levels such as:
\begin{enumerate}
\item Critical Control data that consists of telemetry and flight commands (highest priority and highest level of protection);
\item Mission data comprising mission specific data and instructions (requires confidentiality and integrity);
\item Payload data from sensors, images, videos (level of security depends on the application);
\item Operational log data consisting of non real time data that requires basic security \cite{ref71}.
\end{enumerate}

Beyond data sensitivity, UAV security must be extended across other layers of communication nodes. As highlighted by the authors of \cite{ref72}, threats can arise from other sources such as underlying hardware, software, and sensing components. At the hardware level, UAVs can be exposed to tampering and capturing of drone side-channel attacks (analyzing indirect physical or behavioral data like timing variations to carry out tasks and power consumption patterns). At the software level, unwanted firmware updates and different application codes that allow the unwanted injection of malware can affect the mission. At the sensor level, sensor manipulation, spoofing (GPS) attacks can occur. At the communication level, the wireless channel becomes the targeted component that becomes susceptible to jamming, replay attacks, and manipulation of protocol manipulation (attacks that exploit weaknesses in protocols that are used between UAV and GCS).

Although UAV security study is vast and needs to be enforced across all the above mentioned areas, UAV systems emphasize Physical Layer Security (PLS) because UAVs operate in open and exposed wireless environments that make them vulnerable to eavesdropping and jamming. By establishing a full-fledged solution applicable to UAVs, the focus is not only on the security of data transmission, but also on improving network coverage.

\subsection{Jamming and Eavesdropping and their Mitigation}
\subsubsection{Jamming and Eavesdropping}
Jamming is an active attack in which a hostile node transmits a signal intended to disrupt communications between a UAV and legitimate terrestrial UEs or fellow UAVs. It degrades the Signal-to-Interference-Plus-Noise Ratio (SINR) to an extremely unreliable level. The following types of jamming attacks can be seen:
\begin{enumerate}
    \item Constant jamming, where a constant signal is transmitted to disrupt channel communication; 
    \item Reactive jamming, where a disruptive signal is transmitted only after transmission of the original signal starts. This makes the hostile node harder to detect; 
    \item Deceptive jamming, where a hostile node pretends to be the authorized node and transmits interfering signals \cite{ref73,ref74}. 
\end{enumerate}

Another type of attack, which is passive in nature, is eavesdropping. These types of attacks aim to acquire sensitive information, making the UAVs vulnerable to such attacks, as UAVs have communication channels in the open air, dominant LoS, and pre-planned flight paths \cite{ref75}. Jamming causes  data packet losses, delays in transmission, and C2 link stability issues, while eavesdropping aims to disrupt data confidentiality and integrity \cite{ref73,ref74,ref75}. 

\subsubsection{Mitigation}
To address these attacks and growing threats, a variety of strategies have been proposed and are as follows:
\begin{enumerate}
    \item Cooperative Jamming: In this approach, a swarm of drones (friendly) usually jams the opponents (hostile node) and maintains a secure link with the main channel. This friendly swarm of drones constantly emits an artificial noise signal towards the hostile node, maintaining secrecy and secrecy outage probability (SOP), the probability that the instantaneous secrecy capacity of a communication link falls below a predefined target secrecy rate \cite{ref76}; 
    \item UAV-enabled Trajectory and Path Planning: It consists of two UAVs, one acting as a data transmitter and the other as a jammer. The joint coordination of these helps in the optimization of the trajectory and power, while maximizing the rate of secrecy. Further improvements can be made using reinforcement learning (RL), where an RIS can be used for beamforming \cite{ref77,ref78}; 
    \item Spectrum Allocation and Frequency Hopping: in this approach, UAVs can coordinate for a channel selection in the spectrum; also, frequency hopping develops resistance against these attacks \cite{ref79,ref80}. 

\end{enumerate}

\subsection{Spectrum Allocation and Interference}
UAVs require a dominant LoS when present at high altitudes; UAVs in shared spaces can cause co-channel interference (CCI) (CCI occurs when two or more transmitters use the same frequency channel and signals carried by these channels overlap), which can cause congestion. This can be more challenging if hostile nodes are present, as they can introduce jamming, as flight paths are pre-defined and predictable. To address these issues at the spectrum level, the following mitigation methods have been proposed:
\begin{enumerate}
    \item Multi-Objective Soft Decision Strategy Deep Q-Network (MOSDS-DQN), an RL technique for inter-cell scheduling (allocation of resources like frequency and time-slots) in a multi-cell structure, which helps in reducing UAVs’ interference with each other’s cells. This also results in an increase in throughput by 80 \% \cite{ref81}.
    \item Multi-agent RL, which can be used by UAVs to learn spectral usages and avoid frequencies that often cause jamming \cite{ref82}. 
    \item A 2-stage approach for interference management in inter-cellular UAVs, where a Genetic Algorithm (GA)  that assigns Orthogonal Frequency Division Multiple Access (OFDMA) to communication links between the UAVs and UEs followed by power level organization is applied. This approach achieves 98\% of the objective of reducing interference, followed by low computation times \cite{ref83}. 
    \item Frequency Hopping Spread Spectrum (FHSS), which can be used in UAV coordination and UAV-UE communication, where UEs can switch between different frequencies depending on the priorities, as it makes it difficult for potential jammers to detect the hopping pattern and priorities \cite{ref84}.
\end{enumerate}

\subsection{Trajectory Optimization}
UAVs can dynamically place themselves on a pre-defined flight path (3-D Mobility), which can improve coverage and QoS. PLS can also be elevated more with the help of flight optimization, as the latter not only includes defining paths where power consumption is minimum, but also avoidance of areas that are prone to a high probability of interference from hostile nodes, which minimizes the exposure to eavesdroppers and jammers. Other ideas which can be used for optimization of paths are as follows: 
\begin{enumerate}
    \item  The trajectory can be optimized based on zones, which can offer the maximum secrecy rate along with energy efficiency and reduction in interference \cite{ref85}.
    \item Selection of paths that are closer to legitimate UEs, predicted using Channel State Information (CSI).
    \item Addition of decoy/ dummy flight paths to confuse potential jammers and eavesdroppers, with the addition of uncertainties in the communication links \cite{ref86,ref87}.
\end{enumerate}

UAVs, as explained in earlier sections, are used as different types of communication nodes. Traditional network infrastructures heavily rely on fixed strategies, and traditional models such as LTE/4G and 5G network architectures are generally static and centrally controlled. The adopted methods are inefficient for UAVs, as these (UAV) networks tend to be dynamic. To improve efficiency, the integration of UAVs is being studied with different technologies, and some of them are discussed in this section as a part of evolving technologies that have the potential to outperform and highlight important factors that improve UAV performance parameters. The applications discussed below range from swarm intelligence to next generation network systems. 
\begin{enumerate}
    \item Advancements in swarm robotics to operate in non-centralized and self-operating environments. Integration of multiple UAVs with a framework like Robot Operating System 2-Data Distribution Service (ROS2-DDS). ROS2 is a paradigm for building robot applications, and DDS is the protocol that enables communication between nodes in ROS. The authors in \cite{ref88} propose a multilayer strategy that incorporates RL in the system for a more robust control, showing dynamic performance of the proposed method with up to 20 UAVs used in the simulations. 
    \item Integration of RIS for secure communications is one more evolving technology under consideration. RIS is accompanied by a  dynamic maneuvering, which can be used by the operator of a UAV to avoid interfering paths and form a direct beam towards the targeted devices. This enhances the secrecy capacity as demonstrated by authors of \cite{ref89}, which uses the RIS approach along with trajectory optimization, phase shifts, and also improves the overall throughput by 25 \%.
    \item UAV integration with 6G and AI is gaining popularity, as integration with AI can enable UAVs to execute applications with more optimized factors, such as the ability to calculate real-time parameters like trajectory coordination, jamming, alignment of beam for more accurate transmissions along their accommodation in Ultra-wide band channels and ISAC.
\end{enumerate}
\FloatBarrier
\begin{figure*}
    \centering
    \includegraphics[width=1.0\linewidth]{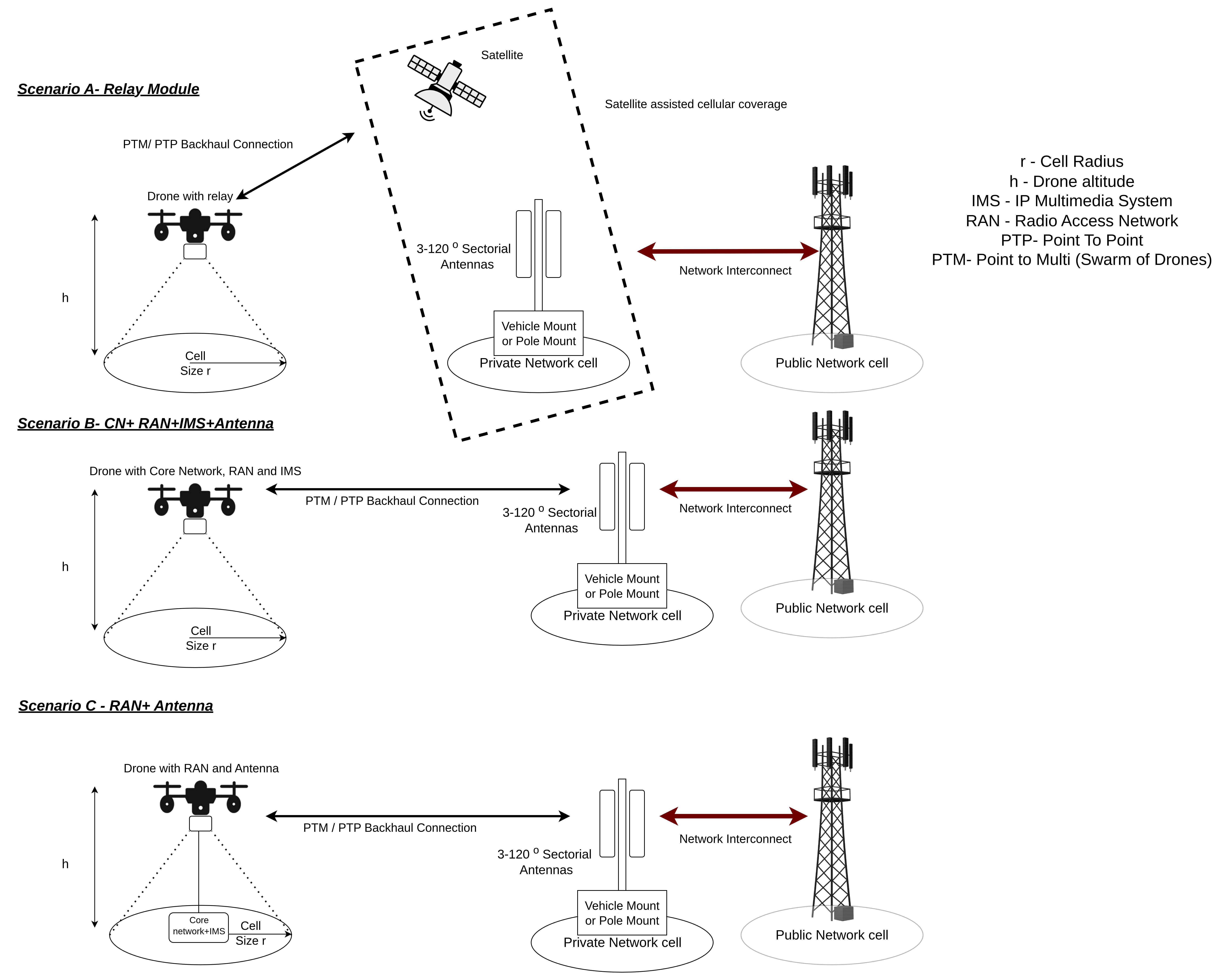}
    \caption{UAV-NIB example configurations}
    \label{fig:12}
\end{figure*}

\section{UAV as a NIB}
The communication landscape is experiencing a paradigm shift towards the integration of UAVs with 5G systems. We present one such innovative approach for disaster management and emergencies-- a NIB on a UAV. A NIB is a portable communication system comprising all wireless components, such as the core network (CN), radio access network (RAN), and IP multimedia subsystem (IMS) for rapid deployment of standalone cellular networks. The aim is to establish a resilient communication system in areas where the deployment of traditional 5G communication infrastructure is not feasible. By combining a NIB with a UAV, we ensure cellular coverage, low-latency data transmission, and real-time connectivity. A NIB is a miniature system that is portable and capable of establishing a whole network without the need for a traditional communication network. Other features include core capabilities and support for multiple standards, low-latency applications, and low power consumption \cite{nib1,nib2}. Some examples of configurations for UAV as a NIB are illustrated in Figure \ref{fig:12} that illustrates: 
\begin{enumerate}
\item  A relay mode, where the UAV operates as a flying mobile relay to extend coverage towards a private network. 
\item  A core network and antenna mode, in which the UAV carries radio access, antenna, and selected core network functions to form an aerial small cell.
\item A RAN-only mode, where the UAV hosts radio and antenna components while core network functions remain on the ground. 
\end{enumerate}
In all cases, the UAV establishes a wireless backhaul link to a public or private communication network infrastructure.

\begin{figure*}
    \centering
    \includegraphics[width=0.8\linewidth]{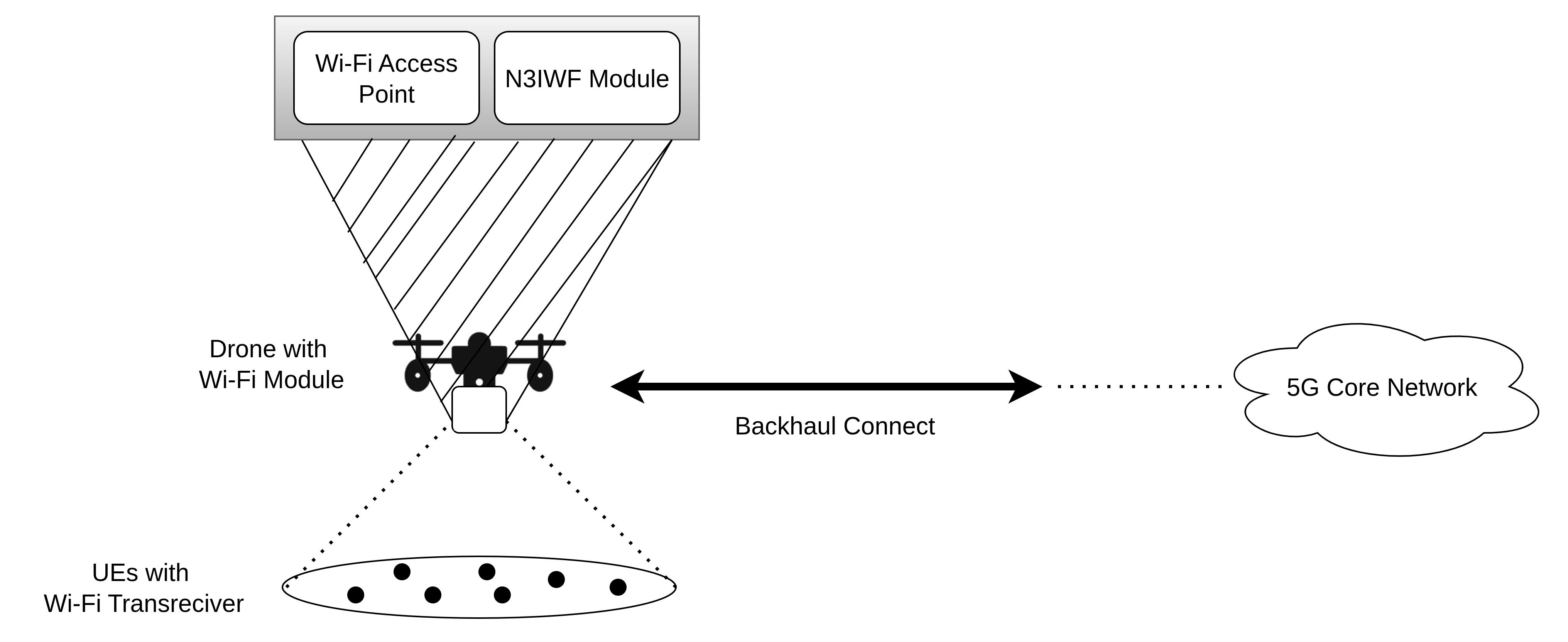}
    \caption{UAV- N3WIF Configuration}
    \label{fig:13}
\end{figure*} 

Many deployment scenarios are possible. Each scenario in the figure depicts different deployment strategies of NIB elements, which are as follows:
\begin{enumerate}
    \item Scenario A uses a satellite connection, making it ideal for coverage extension. The UAV is accompanied by a relay and an antenna module; this setup is not entirely autonomous and relies heavily on a backhaul connection. The latency of this setup is moderate to high and dependent on the backhaul.
    \item In Scenario B, a UAV is equipped with the entire NIB, offering full autonomy and ultra-low latency conditions best suited for isolated deployments. This scenario is limited due to the addition of weight to the payload, which reduces the flight time.
    \item Scenario C is a RAN-only configuration in the payload with CN and IMS as parts of the terrestrial network; this provides a chance to optimize the payload and autonomy, offering scalability. It requires reliable backhaul connectivity and offers a moderate latency in different applications.
\end{enumerate}

Effective and efficient deployment of this configuration requires an understanding of air-ground channels. Channels are often dynamic and prone to multi-path fading, Doppler effects, shadowing, and interference. A constant dominant LoS for persistent communication is needed. Accurate channel models that are 3D in nature are needed, which highly depend on UAV altitude, antenna type, orientation, and hovering capabilities along with surrounding areas (urban, rural, or town) as it determines the wireless signal propagation without interference of RF waves. Awareness of backhaul connectivity is also required. 

The deployment of a UAV as a NIB must comply with the frameworks developed by international bodies such as 3GPP, ITU, and ETSI. All aviation and communication-related protocols, such as spectrum allocation, resource management, and channel models, must be implemented effectively. UAVs with other payloads (NIBs and other devices) will be thoroughly classified. The communication of the configuration will be checked to find if it falls under  Beyond Visual LoS (BVLOS) or  Visual LoS (VLOS), and the safety compliance will be recognized according to the applications. Data protection and lawful interception policies apply to deployed NIBs operating in public service modes, especially if user data is relayed through non-standard infrastructures during emergencies.

Traditional networks are often stable, equipped with high-throughput components, but are inaccessible in cases of emergencies, military applications, and disasters. In contrast, UAV-mounted NIBs can be dynamically deployed as they require minimal on-ground support. They can provide real-time network coverage, multiple standards connectivity (5G, 4G, etc.), and are efficient in energy consumption. They are ideal for temporary scenarios and are far more cost-effective than traditional networks. To summarize, UAV as a NIB is a novel solution for rapid and efficient deployment and extension of the terrestrial network.

One additional deployment scenario, which can be 3GPP compliant, is the integration of Wi-Fi technology via the Non-3GPP Interworking Function (N3WIF). In this architecture, as illustrated in Figure \ref{fig:13}, the UAV acts as an access point (Wi-Fi based), allowing UEs with standard Wi-Fi transceivers to connect with the core network. The N3IWF module is responsible for establishing secure Internet Protocol Security (IPsec) connections to enable session, mobility, and resource management. This setup eliminates the necessity of a traditional gNB between the drone and the core network. This leads to a reduction in power consumption, complexity, and size. It offers flexible deployment during natural disasters, emergencies, and temporary coverage.

\section{Future works}
The current literature reveals many futuristic and critical paths for investigation in UAV-assisted communication. Despite progress in aerial networking, solutions often assume ideal conditions and remain limited to addressing stringent conditions which are mentioned in this section. Future research therefore must focus on bridging the gaps such as realistic contested spectrum environments, dynamic topology changes, limited onboard resources and the need for rapid and autonomous deployment. Some of these promising research directions are summarized in the following.
\begin{enumerate}
 \item  Resilience and Fault Tolerance
While current models focus on software and signal-based anomalies, future research will need to expand real-time monitoring to encompass hardware-specific failures. Specifically, evaluating performance against hardware-induced faults that is (e.g., motor seizure) essential for safety-critical deployments. Furthermore, there is a need to develop fault-tolerant control strategies that allow UAVs to maintain stability or perform safe landings after an event (fault/ hardware or software failure) detection \cite{fut1}.

\item Advanced Resource and Trajectory Optimization
The integration of UAVs into Computing Power Networks (UAV-CPNs) introduces complex energy constraints. Future work should refine jointly optimizing UAV altitude and transmit power specifically under hybrid energy architectures involving fuel cells and batteries \cite{fut2}. Additionally, in IoT data collection, more sophisticated deep reinforcement learning (DRL) frameworks, such as rainbow learning, could be explored to jointly optimize trajectory, receive beamforming, and ground node power allocation in-order to maximize data volume \cite{fut3}.

\item 6G and Non-Terrestrial Network (NTN) integration
As networks transition towards 6G, the role of UAVs as aerial base stations require more robust backhaul solutions. Investigating rotary-wing UAVs equipped with LEO satellite backhaul (e.g., Star-link) for emergency connectivity in rural or compromised environments remains a priority \cite{fut4}. To maintain these links, future research must address high doppler shifts through adaptive beamforming frameworks using "Deep Q-Learning" to align communications with fast-moving LEO satellites (speed?)\cite{fut5}.

\item Mobility Management and Task Offloading- 
Managing connectivity in dynamic environments remains a challenge. Future studies should move beyond signal-strength-based handovers to explore overcoming the limitations of traditional handovers for drone base stations (DBSs) in ubiquitous 6G networks \cite{fut6}. For computational efficiency, the extension of latency-aware algorithms (such as LLDDPG) to handle "diverse MEC nodes and varying UAV image data sizes" will be vital for real-time surveillance and disaster response applications \cite{fut7}.
\end{enumerate}

\section{Conclusions}
This survey explored UAVs in different useful forms of communication roles, such as UE, gNB, relay, and RIS. Diverse applications and conceptual frameworks were explored. Integration with AI and physical layer security (PLS) were also addressed. Special attention was paid to the power source and its impact on the communication payload. The key PLS mechanisms required to safeguard UAV communications within a network were reviewed. Evolving technologies were explored, along with future works and problem statements. The UAV-NIB configuration as a substitute for traditional networks is proposed. The synergy of these technologies positions UAVs as a critical enabler of an intelligent, resilient, and adaptable communication system for next-generation networks.

\section*{Acknowledgment}
\begin{center}
The authors would like to express their sincere gratitude to Shri Pranav Jha and for his valuable guidance and insightful inputs to this work and Mr. Lakhwinder Singh Mamak for his insights during the early discussions.
\end{center}

\bibliographystyle{IEEEtran}  
\bibliography{references}     

\end{document}